\documentclass[
prb,twocolumn,
nofootinbib,
superscriptaddress,
amsmath,amssymb,
aps,floatfix
]{revtex4-2}
\usepackage{graphicx}
\usepackage{xcolor}
\usepackage{bm}
\usepackage{newtxtext}
\usepackage{dsfont}
\usepackage[
colorlinks=true,
urlcolor=blue,
linkcolor=blue,
citecolor=blue
]{hyperref}
\usepackage[capitalize]{cleveref}
\usepackage[english]{babel}
\usepackage[autostyle, english = american]{csquotes}
\MakeOuterQuote{"}
\usepackage[mathlines]{lineno}
\usepackage{braket}
\usepackage{floatrow}
\usepackage[caption=false]{subfig}
\usepackage[export]{adjustbox}
\graphicspath{{./}}

\def \tr{\operatorname{tr}}
\def \sgn{\operatorname{sgn}}

\def \Imag{\operatorname{Im}}
\def \Real{\operatorname{Re}}
\def \U{\mathrm{U}}
\def \SU{\mathrm{SU}}

\begin{document}

\title{Robust spin pseudogap and spin-charge separation in the $\sigma t$-$J$ model}

\author{Zheng-Yuan Yue}
\thanks{These authors contribute equally.}
\affiliation{Department of Physics, The Chinese University of Hong Kong, Sha Tin, New Territories, Hong Kong, China}

\author{Jia-Xin Zhang}
\thanks{These authors contribute equally.}
\affiliation{French American Center for Theoretical Science, CNRS, KITP, Santa Barbara, California 93106-4030, USA}
\affiliation{Kavli Institute for Theoretical Physics, University of California, Santa Barbara, California 93106-4030, USA}

\author{Zheng-Cheng Gu}
\email{zcgu@phy.cuhk.edu.hk}
\affiliation{Department of Physics, The Chinese University of Hong Kong, Sha Tin, New Territories, Hong Kong, China}
\affiliation{The State Key Laboratory of Quantum Information Technologies and Materials, The Chinese University of Hong Kong, Shatin, New Territories, Hong Kong, China}

\date{\today}

\begin{abstract}
    The $\sigma t$-$J$ model, obtained from the usual $t$-$J$ model by flipping the sign of the spin-down hopping term, has been proposed to eliminate the strong interference between doped holes and the spin background, known as the phase string effect.
    In this work, we investigate the finite-temperature properties of the $\sigma t$-$J$ model using infinite projected entangled-pair state (iPEPS) algorithms, in comparison with the $t$-$J$ model and its easy-plane variants.
    We show that these models share similar spin-pseudogap thermodynamics, with a maximum in the spin susceptibility at $T^* \sim J$ and a broad specific-heat peak.
    Tensor network renormalization (TNR) further provides strong evidences for a Berezinskii-Kosterlitz-Thouless (BKT) transition in the spin sector of the $\sigma t$-$J$ model at a lower temperature $T_\text{BKT} < T^*$ due to its reduced $\mathrm{U}(1)$ spin-rotation symmetry.
    Crucially, in contrast to the $t$-$J$ model and to its easy-plane variants that also exhibit BKT transitions, these signatures remain robust upon doping: the doped holes preserve and even enhance the antiferromagnetic correlations in the $xy$ spin plane, leading to a weak doping dependence of both $T^*$ and $T_\text{BKT}$.
    Finally, we develop a slave-fermion mean-field theory for the $\sigma t$-$J$ model, whose projective symmetry group (PSG) is selected based on numerically determined hopping and pairing correlations, and show that it explains the robust spin pseudogap upon doping.
\end{abstract}

\maketitle

\section{Introduction}

Doped Mott antiferromagnets are a prototypical setting for studying the interplay between the spin and charge degrees of freedom generated by strong electron correlation \cite{lee2006doping,keimer2015from}.
Their best-known experimental realization is the family of cuprate high-temperature superconductors, whose parent compounds are antiferromagnetic Mott insulators.
Upon doping, superconductivity emerges at low temperature, while above the superconducting transition temperature the underdoped compounds enter a pseudogap regime \cite{timusk1999pseudogap}. This regime shows up in many experimental probes; for example, nuclear magnetic resonance detects a suppression of low-energy spin excitations \cite{warren1989cu,alloul1989nmr}, and angle-resolved photoemission spectroscopy (ARPES) reveals a gaplike depletion of the low-energy single-particle spectral weight \cite{ding1996spectroscopic,loeser1996excitation}.
The same regime supports a host of competing, intertwined collective-mode fluctuations, which together produce an exceptionally rich variety of experimental signatures \cite{keimer2015from,fradkin2015colloquium,proust2019remarkable,agterberg2020physics}.

The complexity of these emergent phenomena, however, is also a major obstacle to understanding the underlying correlation effects.
For instance, the superconducting order developing at low temperature conceals the intrinsic ground state of the pseudogap regime, so that even its most basic character, metallic or insulating, cannot be probed directly \cite{zhang2023crossover}.
Likewise, the intertwined order fluctuations above the superconducting transition temperature have left the origin of the pseudogap in long-standing debate \cite{keimer2015from,fradkin2015colloquium,agterberg2020physics}.
The one-band $t$-$J$ model \cite{zhang1988effective,lee2006doping} is widely believed to capture the essential physics of these materials; its Hamiltonian is $H = H_t + H_J$, where
\begin{equation}
    \begin{aligned}
        H_t & = -t \sum_{\braket{ij}} (
        \tilde{c}^\dagger_{i\uparrow} \tilde{c}_{j\uparrow}
        + \tilde{c}^\dagger_{i\downarrow} \tilde{c}_{j\downarrow}
        ) + h.c.,
        \\
        H_J & = J \sum_{\braket{ij}} \left(
        \mathbf{S}_i \cdot \mathbf{S}_j - \frac{1}{4} n_i n_j
        \right).
    \end{aligned}
    \label{eq:tJ-hamiltonian}
\end{equation}
Here $\tilde{c}_{i\sigma}$ ($\sigma \in \{\uparrow, \downarrow\}$) is the electron operator projected onto the subspace without double occupancy, $\mathbf{S}_i = \sum_{\alpha \beta} c^\dagger_{i\alpha} \boldsymbol{\sigma}_{\alpha \beta} c_{i\beta} / 2$ is the spin-1/2 operator, and $n_i = \sum_\alpha c^\dagger_{i\alpha} c_{i\alpha}$ is the electron number operator.
The projection encodes the strong on-site repulsion of the Mott insulator directly in the restricted Hilbert space. Within this space the hopping term $H_t$ acts, while the superexchange term $H_J$ supports an antiferromagnetic spin background.

Despite its simple form, the $t$-$J$ model is notoriously difficult to solve.
Numerically, large-scale density matrix renormalization group (DMRG) and tensor network studies of its ground state reveal a delicate competition between charge-density-wave (stripe) and superconducting orders, whose balance is sensitive to model details such as the next-nearest-neighbor hopping \cite{corboz2011stripes,corboz2014competing,jiang2021ground,gong2021robust,li2021study,lu2024sign}.
Analytically, despite a broad spectrum of theoretical approaches \cite{lee2006doping,lee2014amperean,agterberg2020physics,schmittrink1988spectral,kane1989motion}, a comprehensive description of all these complex phenomena remains a challenge.
Such richness makes it hard to disentangle the origin of each phenomenon: whether the different orders cooperate, compete, or are merely by-products of one another.

A natural strategy is therefore to look for a closely related companion model of the doped Mott antiferromagnet.
Such a model should be simpler than the $t$-$J$ model, with fewer competing orders and a well-controlled effective description, while still sharing similar characteristic correlation-driven phenomena.
Comparing the two then provides a controlled way to dissect the intricate interplay between the charge and spin degrees of freedom in doped Mott antiferromagnets.
In this work we study such a companion system, the $\sigma t$-$J$ model \cite{lu2024sign,zheng2025revealing}, obtained from the $t$-$J$ model by simply replacing $H_t$ with
\begin{equation}
    H_{\sigma t} = -t \sum_{\braket{ij}} (
    \tilde{c}^\dagger_{i\uparrow} \tilde{c}_{j\uparrow}
    - \tilde{c}^\dagger_{i\downarrow} \tilde{c}_{j\downarrow}
    ) + h.c.
    \label{eq:stJ-hamiltonian}
\end{equation}
This sign flip of the hopping for spin-down electrons alters the coupling between the mobile charge and the local spins, while leaving the no-double-occupancy constraint and the superexchange $H_J$ untouched \cite{lu2024sign}.
It turns out that this apparently minor change greatly reduces the complex interference between the doped holes and the antiferromagnetic spin background.
We will briefly discuss the physical picture behind this construction at the end of this work in \cref{sec:conclusion}.
The spin-dependent hopping singles out the spin $z$ axis, reducing the spin rotation symmetry from $\SU(2)$ to $\U(1)$ (see proof in Appendix \ref{app:stJ}).
Zero-temperature DMRG \cite{lu2024sign} and iPEPS \cite{zheng2025revealing} studies have shown that the resulting physics is indeed much simpler.
At half filling, the parent state remains the conventional antiferromagnetic Mott insulator. Upon doping, the ground state becomes a Fermi-liquid-like state of holes that coexists with antiferromagnetic long-range order in the $xy$ spin plane, with no additional charge order.

However, to provide more insight for the pseudogap phenomenology, a finite-temperature study is more relevant.
The central object encoding thermal properties is the Gibbs state operator $\rho(\beta) \propto e^{-\beta H}$.
On cylinder geometries, one can use the tangent space tensor renormalization group (tanTRG) algorithm \cite{li2023tangent} to obtain $\rho(\beta)$ as a matrix product operator, which is based on the time-dependent variational principle (TDVP) \cite{haegeman2011time} in imaginary time and has been applied to the $t$-$J$ model and the related Hubbard model \cite{li2023tangent,qu2024phase,li2026thermal,li2026fluctuating}.
Recently, tensor network renormalization (TNR) of the partition function $Z = \tr \rho(\beta)$ has been developed as a direct probe of thermal phase transitions \cite{ueda2026global}, requiring $Z$ to be represented as an infinite 2D tensor network with explicit translation symmetry.
Therefore, to enable TNR diagnostics, we shall instead employ finite-temperature iPEPS and iPEPO algorithms \cite{czarnik2012projected,czarnik2014fermionic,dziarmaga2021time,zhang2026finite} working directly in the thermodynamic limit.

In this work, we find that the $\sigma t$-$J$ model shares the spin-pseudogap thermodynamic features of the $t$-$J$ model: the uniform spin susceptibility develops a maximum at a scale $T^*$, accompanied by a broad specific-heat peak.
The key difference is how these signatures respond to doping. In the $\sigma t$-$J$ model, doped holes preserve and even enhance antiferromagnetic correlations in the $xy$ spin plane, and $T^*$ remains of order $J$ with only a weak doping dependence up to $\delta \approx 0.20$.
Tensor network renormalization further gives strong evidence for a Berezinskii-Kosterlitz-Thouless (BKT) transition \cite{berezinskii1971destruction,berezinskii1972destruction,kosterlitz1973ordering} into a phase with quasi-long-range antiferromagnetic order in the $xy$ plane at a temperature $T_\text{BKT} < T^*$ that also stays nearly constant over a broad range of intermediate doping.
This two-scale structure, in which a spin pseudogap forms at $T^*$ well above the magnetic BKT transition $T_\text{BKT}$, is reminiscent of the pseudogap phenomenology of the underdoped cuprates \cite{alloul1989nmr,timusk1999pseudogap}.
By comparing with the original $t$-$J$ model and with its easy-plane variants, which reduce the spin rotation symmetry while keeping the original strong hole-spin interference, we show that the robustness against doping cannot be attributed to the easy-plane spin response alone; it relies on the altered hole-spin coupling \cref{eq:stJ-hamiltonian} of the $\sigma t$-$J$ model.

From a theoretical standpoint, the reduced hole-spin interference makes the $\sigma t$-$J$ model a natural setting for a slave-particle description based on spin-charge separation.
Since the doped holes form a Fermi-liquid-like component while the spins retain robust antiferromagnetic correlations, we use a slave-fermion representation with fermionic holons and Schwinger-boson spinons \cite{jayaprakash1989mean,yoshioka1989slave,arovas1988functional,read1991large}.
The simplicity of the $\sigma t$-$J$ model then makes a mean-field treatment already quite informative, but the projective symmetry group (PSG) classification of symmetric mean-field ans\"atze still admits many inequivalent classes \cite{wen2002quantum,savary2017quantum,wang2006spin,yang2016schwinger}, making symmetry alone insufficient to pin down the effective theory.
Here we use the tensor network results as an additional guide: the zero-temperature correlations observed numerically strongly constrain the PSG-allowed parameter space. This lets us construct a specific slave-fermion mean-field theory for the $\sigma t$-$J$ model.
The resulting mean-field theory describes the ground state faithfully, including the doping-enhanced in-plane antiferromagnetism, and it also captures the finite-temperature features found numerically, reproducing the spin pseudogap at $T^*$ and providing a natural spin-sector interpretation of the BKT transition.

The rest of this paper is organized as follows.
\cref{sec:algorithm} briefly reviews the finite-temperature iPEPS and loop-TNR algorithms used in this work.
\cref{sec:num-result} presents the finite-temperature numerical results for the $\sigma t$-$J$ model and contrasts them with the original $t$-$J$ model and its easy-plane variants, while Appendix \ref{app:xxz} benchmarks the loop-TNR diagnostics of BKT transitions using pure spin models.
\cref{sec:mft} develops the PSG-constrained slave-fermion mean-field theory that accounts for the numerical findings, with technical details collected in Appendix \ref{app:mft}.
\cref{sec:conclusion} summarizes the results and discusses open directions.

\section{Finite-temperature tensor network algorithms}
\label{sec:algorithm}

\subsection{Fermionic tensors}
\label{sec:fermionic-tensor}

To deal with the anticommuting nature of fermionic degrees of freedom, we adopt the formalism of \emph{fermionic tensors} \cite{bultinck2017fermionic,mortier2025fermionic}.
The basic idea is to replace ordinary tensor legs by $\mathbb{Z}_2$-graded \emph{super vector spaces} $V = V^0 \oplus V^1$, whose basis states carry a fermion parity $|i| = 0,1$.
A fermionic tensor is an element of a tensor product of such spaces and their duals, for example
\begin{equation}
    A = A_{\alpha\beta\gamma}\ket{\alpha}\bra{\beta}\ket{\gamma},
    \label{eq:tensor-example}
\end{equation}
with each index assigned a parity.
All tensors are required to have even parity, meaning that tensor elements vanish unless the total parity of all indices is even.
The fermionic signs enter through the elementary tensor operations.
First, changing the order of two basis states $\ket{i}, \ket{j}$ produces the exchange sign of fermionic modes,
\begin{equation}
    \ket{i}\ket{j}
    \mapsto
    (-1)^{|i||j|}\ket{j}\ket{i},
\end{equation}
Second, each leg is oriented according to whether it represents a vector or a dual vector, and contractions are defined between compatible dual and non-dual spaces.
Graphically, a vector is represented by an outgoing arrow, while a dual vector is represented by an incoming arrow. For example, the tensor $A$ in \cref{eq:tensor-example} is graphically represented as
\begin{equation}
    \begin{matrix}
        \includegraphics[width=0.25\columnwidth]{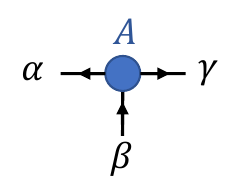}
    \end{matrix}.
\end{equation}
For convenience, in the following we shall draw the arrows only when necessary. The elementary contraction map is the usual evaluation map for a bra followed by a ket,
\begin{equation}
    \mathcal{C} : V^* \otimes V \to \mathbb{C},
    \quad
    \mathcal{C}(\bra{i}\ket{j}) = \braket{i|j} = \delta_{ij}.
    \label{eq:contract}
\end{equation}
If the contracted pair appears in the opposite order, it must first be fermionically reordered,
\begin{equation}
    \mathcal{C}(\ket{i}\bra{j})
    =
    (-1)^{|i||j|} \mathcal{C}(\bra{j}\ket{i})
    =
    (-1)^{|i|}\delta_{ij}.
    \label{eq:super-contract}
\end{equation}
Thus, a ket-before-bra contraction gives the \emph{supertrace} parity factor.
In this manner, one does not need to manually resolve and insert explicit fermionic swap gates \cite{corboz2010simulationstrongly} at every line crossing, which allows a straightforward generalization of all tensor network algorithms to fermionic systems.

\subsection{Obtaining the Gibbs state with Neighborhood Tensor Update}
\label{sec:ntu}

The finite-temperature Gibbs operator $\rho(\beta) \propto e^{-\beta H}$ is obtained from its purification $\psi(\beta)$ \cite{verstraete2004matrix,czarnik2012projected}, which is a state living in an enlarged Hilbert space composed of the original physical space ($p$) and its \emph{dual} copy called the ancilla ($a$).
The Gibbs operator is recovered by tracing out the ancilla spaces,
\begin{equation}
    \rho(\beta) = \tr_a \psi(\beta) \psi^\dagger(\beta).
    \label{eq:purify}
\end{equation}
At infinite temperature $\beta = 0$, the Gibbs operator reduces to the identity operator on the physical spaces.
Its purification $\psi(0)$ is an iPEPS with two "physical legs" at each site and virtual bond dimension $D = 1$.
In this state, each physical site is maximally entangled with its ancilla,
\begin{equation}
    \psi(0) = \bigotimes_i \sum_{s_i} \ket{s_i}_p \bra{s_i}_a,
\end{equation}
where $s_i$ sums over the three basis states (spin-up, spin-down, empty) at site $i$.
The finite-temperature state $\psi(\beta)$ is then obtained by imaginary-time evolution of the physical degrees of freedom,
\begin{equation}
    \psi(\beta)
    \propto U(\beta/2) \psi(0), \quad
    U(\beta / 2) = e^{-\beta H/2},
\end{equation}
so that \cref{eq:purify} gives $\rho(\beta) \propto e^{-\beta H}$.Throughout the following calculations, we assume a checkerboard structure of $\psi(\beta)$ consisting of only two independent tensors, focusing on uniform states without stripe order or other long-range modulations.

When $H = \sum_{\braket{ij}} H_{ij}$ is a nearest-neighbor Hamiltonian, the evolution operator $U(\beta)$ can be Trotter-decomposed into two-site gates $g_{ij} = e^{-\epsilon H_{ij}}$ with a small time step $\epsilon$ acting on nearest-neighbor bonds $(i, j)$.
In this work, we use first-order Trotter decomposition with $\epsilon = 0.005$.
The gate enlarges the dimension of the virtual bond connecting the two updated tensors; the main numerical task is then to truncate this bond back to a chosen finite dimension $D$,
\begin{equation}
    \begin{matrix}
        \includegraphics[width=0.8\columnwidth]{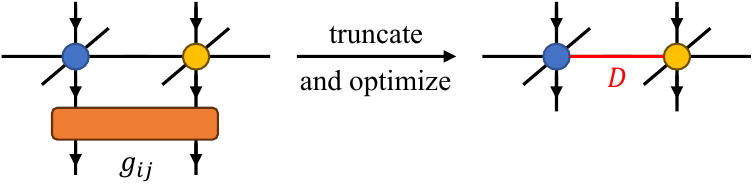}
    \end{matrix}.
\end{equation}
Here we hide the arrows on virtual legs, because they can be chosen arbitrarily.
In this work, the truncation is done with the neighborhood tensor update (NTU) algorithm \cite{dziarmaga2021time,zhang2026finite}, which truncates and optimizes the updated tensors with an environment obtained from their neighboring tensors (here we use the NN+ environment introduced in \cite{king2025beyond}) rather than an approximation of the entire iPEPS.
This makes the update considerably cheaper than a full-update scheme \cite{corboz2014competing} but more accurate than a purely local simple update \cite{jiang2008accurate}.
After reaching the desired temperature $\beta$, we obtain $\psi(\beta)$ in the form of an iPEPS with two "physical legs" at each site and bond dimension $D$. The Gibbs operator $\rho(\beta)$ obtained from \cref{eq:purify} is then an iPEPO with bond dimension $D^2$, which is illustrated as
\begin{equation}
    \begin{matrix}
        \includegraphics[scale=0.5]{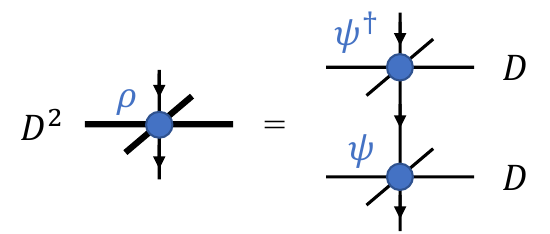}
    \end{matrix}.
\end{equation}
Finally, the expectation values of a physical observable $O$ at inverse temperature $\beta$, given by
\begin{equation}
    \braket{O}(\beta)
    = \frac{\tr[\rho(\beta) O]}{\tr[\rho(\beta)]}
\end{equation}
is measured by contracting the resulting infinite tensor network with the corner transfer matrix renormalization group (CTMRG) algorithm \cite{corboz2014competing}.
The CTMRG environment dimension $\chi$ controls the accuracy of the measurement results.

\subsection{Physical observables at fixed doping}
\label{app:fix-doping}

The doping in the $t$-$J$ model and related systems defined as
\begin{equation}
    \delta = \frac{1}{N} \sum_i \biggl(
    1 - \sum_\sigma \braket{c^\dagger_{i\sigma} c_{i\sigma}}
    \biggr)
\end{equation}
is controlled by adding $-\mu \sum_{i,\sigma} c^\dagger_{i\sigma} c_{i\sigma}$ to the Hamiltonian and tuning the chemical potential $\mu$.
The expectation values of all physical observables (including $\delta$) are naturally functions $O(T,\mu)$ of the temperature $T$ and the chemical potential $\mu$.
If the value of $O$ at a specific $\delta$ is needed, we first scan over a set of chemical potentials $\{\mu_i\}$, obtaining a set $\{O(T,\mu_i)\}$ of the observable and a set $\{\delta(T,\mu_i)\}$ of the doping.
The relation between $O$ and $\delta$ is then obtained by a fourth-order polynomial fit for $\{O(T,\mu_i)\}$ with respect to $\{\delta(T,\mu_i)\}$.
Values of $O$ at a fixed doping are then extracted from the fitted function $O(T,\delta)$.
For quantities derived from more basic observables, such as the specific heat $C_v = \partial E / \partial T$, we first fit the doping dependence of the basic observable (energy $E$ in this case), and then take the numerical derivative.

\subsection{Detecting thermal phase transitions with loop-TNR}
\label{sec:looptnr}

\begin{figure}[tb]
    \centering
    \includegraphics[width=0.98\columnwidth]{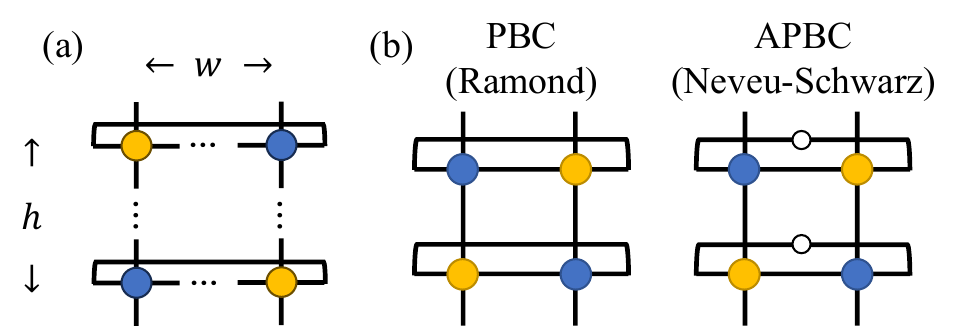}
    \caption{
        (a) A transfer matrix along the vertical direction consisting of $w$ columns and $h$ rows of tensors.
        (b) Fermionic $2 \times 2$ transfer matrices along the vertical direction with PBC (left) and APBC (right) imposed on the horizontal direction. The white circle on the trace in the APBC transfer matrix refers to the fermionic twist.
    }
    \label{fig:transfer-mat}
\end{figure}

After obtaining $\rho(\beta)$ from \cref{eq:purify}, we take the \emph{ordinary} trace over the physical input-output indices on every site. This gives a two-dimensional tensor network representation of the partition function $Z = \tr e^{-\beta H}$ (up to an overall normalization factor), imposing anti-periodic boundary conditions in the imaginary-time direction,
\begin{equation}
    \begin{matrix}
        \includegraphics[width=0.36\columnwidth]{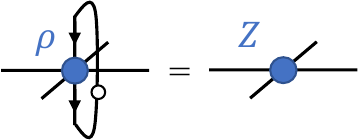}
    \end{matrix}.
    \label{eq:tr-rho}
\end{equation}
We note that a naive contraction of the physical legs following \cref{eq:super-contract} will instead lead to the \emph{supertrace}.
In practice, the unwanted fermionic sign from \cref{eq:super-contract} is cancelled out by inserting a \emph{fermionic twist} \cite{mortier2025fermionic}, i.e., multiplying by $-1$ when summing over tensor elements with odd fermion parity.
Graphically, the twist is represented by the small white circle on the physical leg trace in the left-hand side of \cref{eq:tr-rho}.

The partition-function tensor network can be coarse grained directly using tensor network renormalization (TNR) algorithms.
If the system is near a continuous phase transition, the TNR flow retains scale-invariant information associated with the underlying critical theory.
Here we apply the loop-TNR algorithm \cite{yang2017loop}, which removes short-range entanglement before coarse graining and thereby improves the stability of the RG flow compared with more direct blocking schemes.
To reduce computational cost caused by the large iPEPO bond dimension $D^2$, before tracing over the physical legs of $\rho$, we first truncate the virtual bonds to a smaller dimension $D_\text{pre}$ by inserting a pair of oblique projectors $P_1, P_2$ on each bond that minimize the cost function
\begin{equation}
    \left\Vert
    \begin{matrix}
        \includegraphics[scale=0.5]{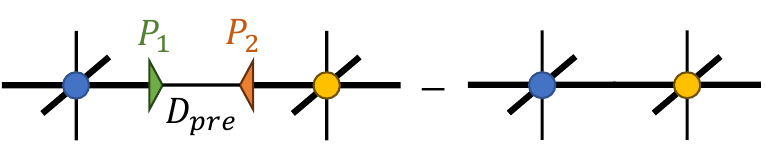}
    \end{matrix}
    \right\Vert^2.
    \label{eq:proj-cost}
\end{equation}

During the loop-TNR flow we monitor the renormalized tensors after successive coarse-graining steps.
If there is a thermal phase transition described by a two-dimensional conformal field theory (CFT), these tensors encode the corresponding universal data.
The CFT spectrum can be extracted from the \emph{transfer matrices} constructed with the renormalized tensors [\cref{fig:transfer-mat}(a)] using the following relation between CFT central charge $c$, scaling dimensions $\Delta_i$, conformal spins $s_i$ and eigenvalues $\lambda_i$ of a transfer matrix \cite{gu2009tensor},
\begin{equation}
    \lambda_i = e^{fN} \exp\left[
        -2\pi \Imag(\tau) \left(\Delta_i - \frac{c}{12}\right)
        + 2\pi i \Real(\tau) s_i
        \right],
\end{equation}
where $\tau$ is the modular parameter of the transfer matrix, $e^f$ is an overall normalization factor, and $N$ is the number of tensors included in the transfer matrix.
$\tau$ and $f$ can be determined by constructing multiple transfer matrices with different geometries \cite{gu2009tensor}.

For fermionic tensor networks, one needs to include all \emph{spin structure} sectors when constructing transfer matrices to extract the complete conformal spectrum \cite{bultinck2017fermionic,bao2019loop}.
When all tensor contractions follow the usual rules of fermionic tensors \cite{mortier2025fermionic}, the resulting transfer matrix has periodic boundary conditions (PBC), yielding the spectrum only in the Ramond (R) sector.
A fermionic twist should be inserted in order to obtain the spectrum in the Neveu-Schwarz (NS) sector with anti-periodic boundary conditions (APBC) [see \cref{fig:transfer-mat}(b)].

\section{Finite-temperature behavior of the \texorpdfstring{$\sigma t$-$J$}{σt-J} model}
\label{sec:num-result}

In this section, we present numerical results from iPEPS simulations.
In both $\sigma t$-$J$ and $t$-$J$ models, we set $t = 2$ and $J = 1$ to compare with a recent finite-temperature iPEPS study of the $t$-$J$ model \cite{zhang2026finite}.
To avoid violation of the Mermin-Wagner theorem \cite{mermin1966absence}, which prevents breaking of continuous symmetries at finite temperature in 2D systems with short-range interactions, we impose the full spin-$\U(1)$ $\times$ charge-$\U(1)$ symmetry of the $\sigma t$-$J$ model, and the spin-$\SU(2)$ $\times$ charge-$\U(1)$ symmetry of the $t$-$J$ model on the tensor network.

Besides the $\sigma t$-$J$ and original $t$-$J$ models, we use two auxiliary comparisons to separate the effects of reduced spin symmetry from those of the strong hole-spin interference.
One is the $t$-XXZ model, which replaces $H_J$ of \cref{eq:tJ-hamiltonian} by
\begin{equation}
    H'_J = \sum_{\braket{ij}} \left[
        J \left(
        S^x_i S^x_j + S^y_i S^y_j - \frac{1}{4} n_i n_j
        \right) + J_z S^z_i S^z_j
        \right].
    \label{eq:txxz-hamiltonian}
\end{equation}
We focus on the $J_z = 0$ limit, referred to as the $t$-XX model, to maximize the easy-plane exchange anisotropy while retaining the original hole-spin interference.
The other is obtained by adding a uniform Zeeman field along the $z$ axis to the $t$-$J$ model,
\begin{equation}
    H^z_h = - h \sum_i S^z_i.
    \label{eq:zeeman-term}
\end{equation}
As long as the uniform magnetization $m_z = \sum_i \braket{S^z_i} / N$ remains small, it effectively induces an easy-plane AFM response in the $xy$ spin plane.
Both variants keep the hopping term $H_t$ unchanged, so the strong hole-spin interference remains.

Because of limited computational resources, all NTU calculations are performed with iPEPS bond dimension $D = 12$ for $\sigma t$-$J$ and $t$-XX, and $D \approx 18$ for $t$-$J$.
The CTMRG environment dimension is $\chi = 64$ unless otherwise stated.
For loop-TNR, we use $D_\text{TNR} = 24$ and $D_\text{pre} = 2 D_\text{TNR} = 48$.
Therefore, the present data should not be regarded as a high-precision determination of the phase diagrams of the Hamiltonians considered here, which would require larger bond dimensions or further algorithmic improvements.
NTU and CTMRG algorithms are implemented in PEPSKit.jl \cite{brehmer2026pepskit}. The loop-TNR algorithm and conformal data calculation are implemented in TNRKit.jl \cite{vanthilt2026practical,vanthilt2026practical-r0.7}. Both packages depend on TensorKit.jl \cite{devos2026tensorkit}, which provides support for fermionic tensors and both Abelian and non-Abelian symmetries.

\subsection{Enhanced antiferromagnetism in the \texorpdfstring{$xy$}{xy}-plane caused by doped holes}
\label{sec:3p-correlator}

\begin{figure}[tb]
    \centering
    \includegraphics[width=0.9\columnwidth]{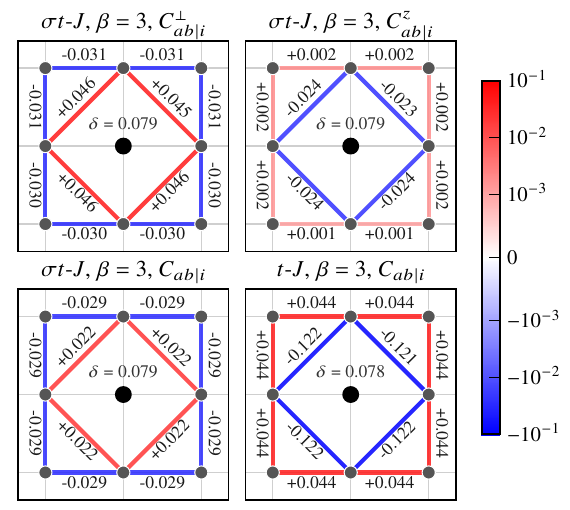}
    \caption{The 3P conditional correlators $C_{ab|i}$ involving NN or NNN spins at temperature $\beta = 3$ and doping $\delta \approx 0.078$ in $\sigma t$-$J$ and $t$-$J$ models. $i$ is the center site, while $(a, b)$ is an NN or NNN bond in the $3 \times 3$ window not overlapping with $i$.}
    \label{fig:stJ-tJ-polaron-config}
\end{figure}

\begin{figure}[tb]
    \centering
    \includegraphics[width=\columnwidth]{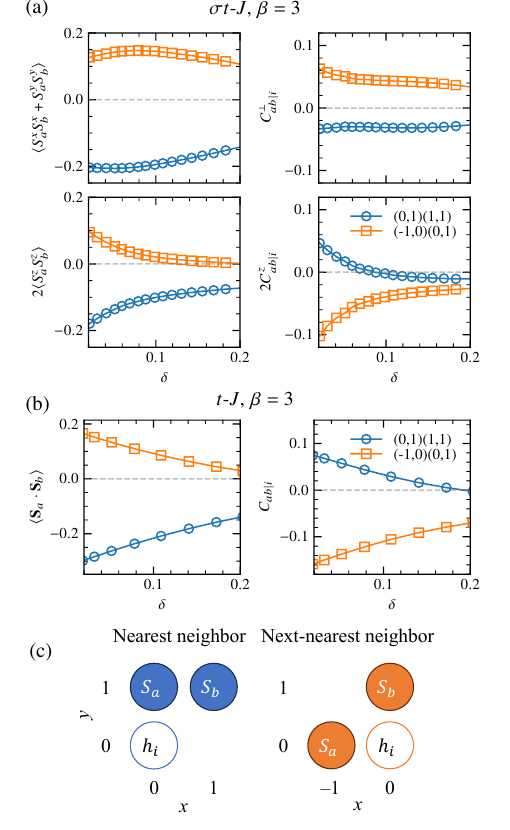}
    \caption{
        The spin correlation $\braket{\mathbf{S}_a \cdot \mathbf{S}_b}$ and the 3P correlator $C_{ab|i}$ as a function of doping at temperature $\beta = 3$ in (a) the $\sigma t$-$J$ model and (b) the $t$-$J$ model.
        (c) The hole is at site $i = (0,0)$, and the two spins are at site $a$ and $b$ specified by the legend.
    }
    \label{fig:stJ-tJ-polaron-doping}
\end{figure}

The effect of doped holes on the spin background can be described by the 3-point (3P) conditional correlators \cite{zhang2026finite} $C_{ab|i} = C^{\perp}_{ab|i} + C^{z}_{ab|i}$, where
\begin{equation}
    \begin{aligned}
        C^{\perp}_{ab|i}
         & = \frac{\braket{B^\perp_{ab} h_i}}{\braket{h_i}} - \braket{B^\perp_{ab}},
        \\
        C^{z}_{ab|i}
         & = \frac{\braket{B^z_{ab} h_i}}{\braket{h_i}} - \braket{B^z_{ab}}.
    \end{aligned}
\end{equation}
Here $h_i = 1 - n_i$ is the hole number operator, $B^\perp_{ab} = S^x_a S^x_b + S^y_a S^y_b$ and $B^z_{ab} = S^z_a S^z_b$ are the $xy$ and the $z$ components of the spin correlation $B_{ab} = \mathbf{S}_a \cdot \mathbf{S}_b$, respectively.
\cref{fig:stJ-tJ-polaron-config} shows the 3P conditional correlator for nearest neighbor (NN) and next-nearest neighbor (NNN) bonds with a $3 \times 3$ window around a hole at the center at temperature $T = 1/3$ and doping $\delta \approx 0.078$, and \cref{fig:stJ-tJ-polaron-doping} shows the doping dependence of the spin correlation and the 3P conditional correlator involving an NN bond and an NNN bond.
For the original $t$-$J$ model, we reproduce the results in \cite{zhang2026finite}: $C_{ab|i}$ is positive for NN bonds and negative for NNN bonds, meaning that the AFM correlation ($\braket{B_{ab}} < 0$) on NN bonds and the FM correlation ($\braket{B_{ab}} > 0$) on NNN bonds are both suppressed.
In the $\sigma t$-$J$ model, the doped holes affect the $xy$ and the $z$ components of the spin correlation differently.
First, similar to the $t$-$J$ model, $C^z_{ab|i}$ is positive for NN bonds and negative for NNN bonds, meaning that the AFM correlations along the $z$ spin direction are suppressed.
However, $C^\perp_{ab|i}$ is negative for NN bonds and positive for NNN bonds. The magnitude of the $xy$ component $|B^\perp_{ab}|$ even slightly increases as doping increases from 0 to $\approx 0.06$.
Therefore, the AFM correlation in the $xy$ spin plane is \emph{enhanced} in the $\sigma t$-$J$ model, which is drastically different from the $t$-$J$ model.
As a result, the spin correlation is dominated by the $xy$ component, with $\braket{B^z_{ab}}$ decreasing faster than $\braket{B^\perp_{ab}}$ as doping increases.
This is also consistent with the larger staggered magnetization in the ground state of the $\sigma t$-$J$ model compared with the $t$-$J$ model \cite{zheng2025revealing}.
The local correlators therefore suggest that reducing the hole-spin interference allows doped holes to preserve, and even enhance, the in-plane antiferromagnetic background.
We next ask how this difference affects finite-temperature bulk thermodynamics, especially in the pseudogap phenomenology previously observed in the original $t$-$J$ model.

\subsection{Robust spin pseudogap upon doping}
\label{sec:pseudogap}

\begin{figure*}[tb]
    \centering
    \includegraphics[width=\columnwidth]{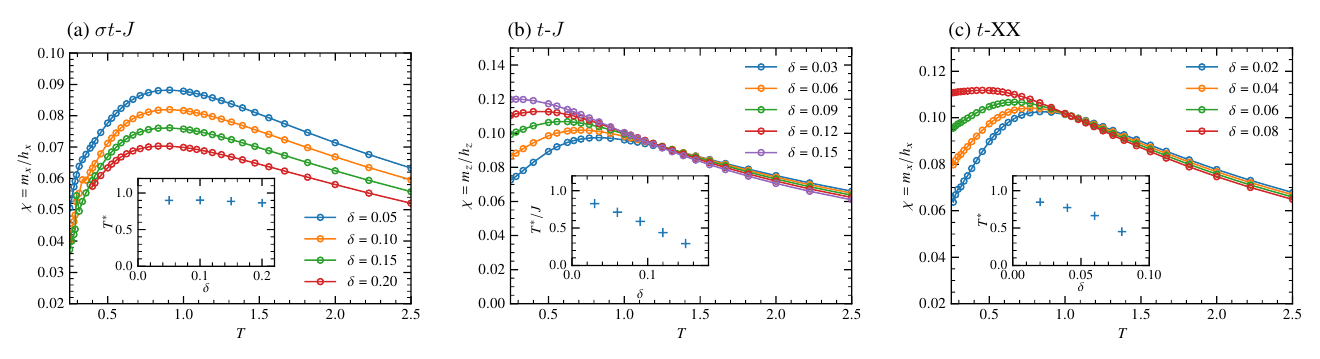}
    \caption{
        Spin susceptibility $\chi$ obtained with a small probe field for (a) $\sigma t$-$J$ model ($D = 12$), (b) $t$-$J$ model ($D = 18$) and (c) $t$-XX model ($D = 12$), all with $t = 2, J = 1$. The field is applied along the $x$ direction for (a,c) and along the $z$ direction for (b). The inset of each panel shows the temperature $T^*$ at which $\chi$ reaches a maximum at a specified doping level.
    }
    \label{fig:mag-chi}
\end{figure*}

\begin{figure*}[tb]
    \centering
    \includegraphics[width=\columnwidth]{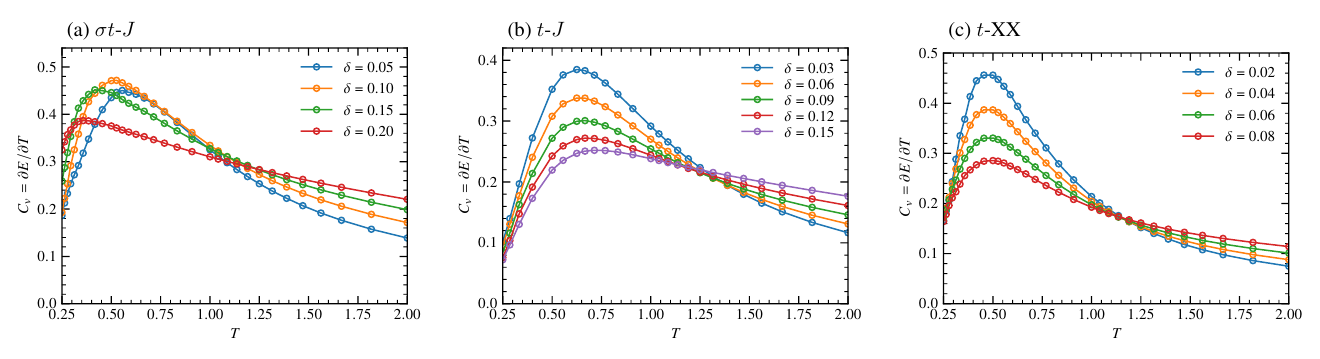}
    \caption{
        Specific heat $C_v = \partial E / \partial T$ for (a) $\sigma t$-$J$ model ($D = 12$), (b) $t$-$J$ model ($D = 18$) and (c) $t$-XX model ($D = 12$), all with $t = 2, J = 1$.
        The energy $E$ excludes the chemical potential terms.
    }
    \label{fig:specheat}
\end{figure*}

We first examine the spin pseudogap, characterized by a maximum in the uniform spin susceptibility occurring at an onset temperature $T^*$ due to the formation of spin singlets, which suppress the alignment of spins with the external magnetic field \cite{zhang2026finite}.
Adding a weak probe field $H_h = - h_a \sum_i S^a_i$, the magnetic susceptibility is estimated from
\begin{equation}
    \chi \approx \frac{m_a}{h_a}, \quad
    m_a = \frac{1}{N} \sum_i \braket{S^a_i}.
    \label{eq:susceptibility}
\end{equation}
For the $\sigma t$-$J$ and $t$-XX models we take $a = x$ to probe the in-plane response, while for the $\SU(2)$-symmetric $t$-$J$ model we take $a = z$.
We set the probe field $h_a = 0.1$, which has been shown to be small enough for the $t$-$J$ model \cite{zhang2026finite}.
As shown by \cref{fig:mag-chi}, all three models exhibit a maximum of $\chi$ at a certain temperature $T^*$.
In this sense, the $\sigma t$-$J$ model retains the same pseudogap phenomenology as the original $t$-$J$ model.
The important difference is the doping dependence of $T^*$.
In both the original $t$-$J$ model and the $t$-XX model, $T^*$ decreases significantly as doping increases \cite{zhang2026finite}.
However, in the $\sigma t$-$J$ model, $T^* \approx 0.9$ has only a weak doping dependence over the studied range $\delta \le 0.20$, with a mild decrease at the largest doping.

The same contrast appears in the specific heat $C_v(T) = \partial E / \partial T$ shown in \cref{fig:specheat}.
Below $T^*$, the formation of short-range spin singlet correlations leads to a broad peak in $C_v$.
In the $\sigma t$-$J$ model, the height and position of the $C_v$ peak remain approximately unchanged over the intermediate-doping range $0.05 \le \delta \le 0.15$, and only at larger $\delta = 0.20$ does the peak shift to lower temperature and become flattened.
In contrast, in the $t$-$J$ and $t$-XX models the $C_v$ peak is quickly flattened as doping increases, consistent with the disruption of spin singlets by the strong hole-spin interference.
Thus, the $\sigma t$-$J$ model is not distinguished by the mere presence of pseudogap-like thermodynamic signatures, but by how robust those signatures remain after doping.

\subsection{Evidence for a Berezinskii-Kosterlitz-Thouless transition of the spin sector at finite doping}
\label{sec:stj-bkt}

\begin{figure}[tb]
    \centering
    \includegraphics[width=\columnwidth]{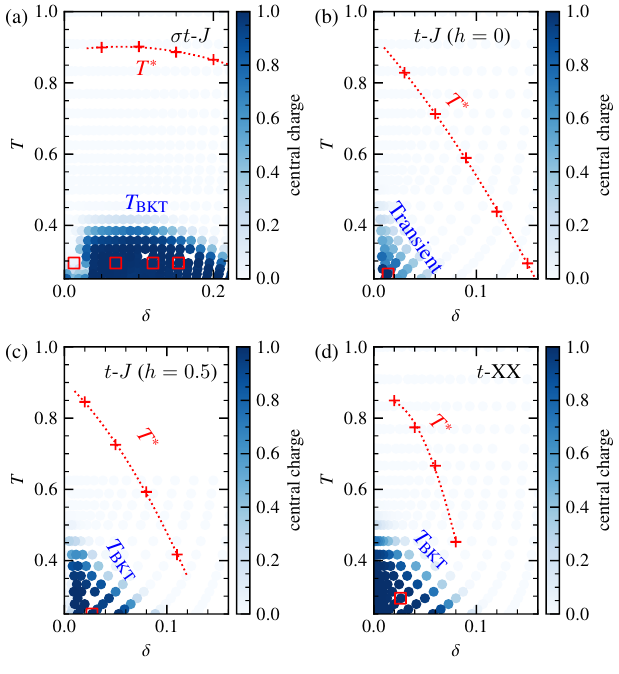}
    \caption{
        Finite-temperature diagram of the central charge at RG step 10 for (a) the $\sigma t$-$J$ model ($D = 12$), (b) the $t$-$J$ model ($D = 18$), (c) the $t$-$J$ model with a uniform magnetic field $h = 0.5$ in $z$ direction ($D = 18$), and (d) the $t$-XX model ($D = 12$). In all cases we set $t = 2$, $J = 1$.
        The CFT spectra of data points marked by a red square are shown in \cref{fig:cft-spec}.
    }
    \label{fig:phases}
\end{figure}

\begin{figure}[tb]
    \centering
    \includegraphics[width=0.72\columnwidth]{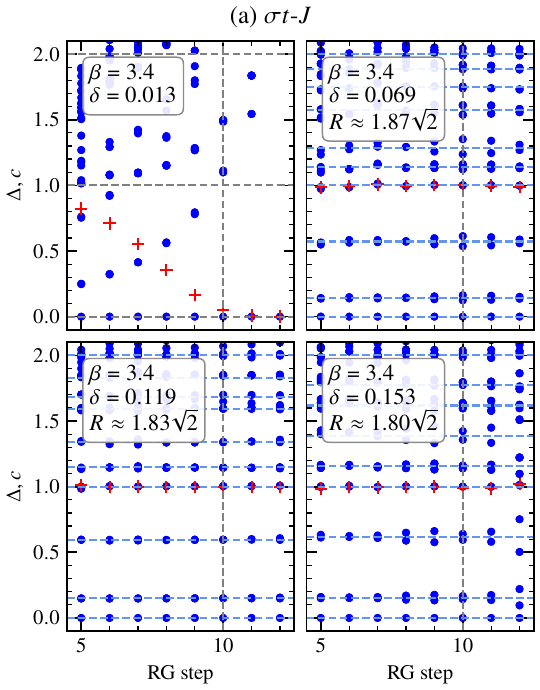}
    \\
    \includegraphics[width=\columnwidth]{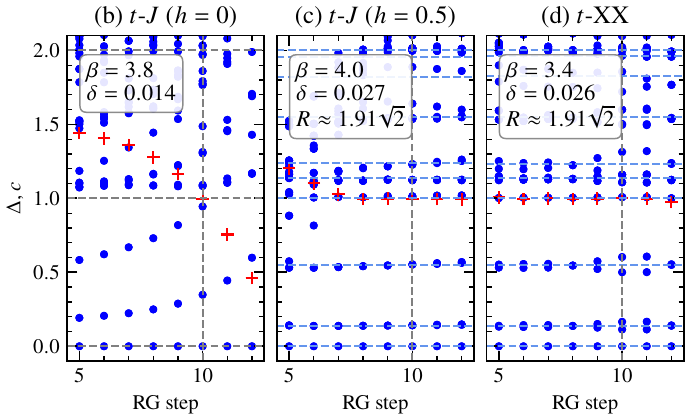}
    \caption{
        CFT central charge $c$ (red crosses) and scaling dimension spectrum $\Delta$ (blue dots, including both NS and R sectors) for data points marked by red squares in \cref{fig:phases}.
        $R$ is the compactification radius obtained from the lowest nonzero $\Delta$ using \cref{eq:boson-cft}.
        The exact compactified boson spectrum is shown by blue dashed lines.
    }
    \label{fig:cft-spec}
\end{figure}

The pseudogap temperature $T^*$ discussed above is a thermodynamic crossover scale associated with short-range spin correlations.
We next ask whether the spin sector undergoes a Berezinskii-Kosterlitz-Thouless (BKT) transition \cite{berezinskii1971destruction,berezinskii1972destruction,kosterlitz1973ordering} at a temperature $T_\text{BKT}$ when the spin rotation symmetry is reduced to $\U(1)$, which can be detected using loop-TNR.

\cref{fig:phases}(a) shows the finite-temperature central-charge diagram of the $\sigma t$-$J$ model.
There is an extended $c \approx 1$ region at low temperature when the doping $\delta \gtrsim 0.04$.
The estimated transition temperature is around $0.33$ in the intermediate-doping regime $0.05 \lesssim \delta \lesssim 0.15$, but starts to decrease at larger doping.
As examined in \cref{fig:cft-spec}(a), for representative points in this $c \approx 1$ region, we observe a relatively stable RG flow even with the huge truncation of the bond dimension from $D^2$ to $D_\text{pre}$. The low-lying scaling dimensions match the compactified-boson CFT, in which the primary fields have scaling dimensions labeled by two non-negative integers $e, m$ \cite{francesco2012conformal},
\begin{equation}
    \Delta_{e,m} = \frac{e^2}{R^2} + \frac{m^2 R^2}{4}
    \label{eq:boson-cft}
\end{equation}
where $R$ is the compactification radius that can be extracted from the lowest nonzero scaling dimension.
These provide strong evidence that the $c \approx 1$ region corresponds to a BKT transition at finite temperature $T_\text{BKT}$.
Although the $\sigma t$-$J$ RG flow is less stable than the purely bosonic XXZ model \cref{eq:xxz-hamiltonian} benchmarked in Appendix \ref{app:xxz} (\cref{fig:xxz-kt}), it is qualitatively distinct from an unstable transient flow, where the spectrum continuously drifts away from the initial values.

For the $t$-$J$ model, if we reduce the spin rotation symmetry to $\U(1)$ by adding a uniform Zeeman field $h = 0.5$ via \cref{eq:zeeman-term}, or by introducing exchange anisotropy in the $t$-XX model, we can also observe a BKT phase [\cref{fig:phases}(c,d)].
Although $h = 0.5$ is strong, the induced uniform magnetization $m_z = N^{-1}\sum_i\braket{S^z_i}$ remains small in the plotted region, with $m_z < 0.08$, so the field mainly reduces the spin symmetry rather than fully polarizing the spins.
However, because of the strong interference caused by the doped holes, $T_\text{BKT}$ remains strongly doping dependent, and the BKT region is restricted to the lower-left corner of the corresponding phase diagrams.
For the original $t$-$J$ model with full $\SU(2)$ spin symmetry, \cref{fig:phases}(b) shows only weak finite-$c$ signals at very small doping.
The representative flow in \cref{fig:cft-spec}(b) is an unstable transient flow: the central charge starts around $1.5$ and then decreases with RG step, similar to the Heisenberg model \cite{ueda2022tensor} (see also Appendix~\ref{app:xxz}), rather than forming a stable plateau.
On the other hand, at small doping, the $\sigma t$-$J$ model also approaches the Heisenberg model, with relatively weak spin rotation symmetry breaking. Therefore, a stable compactified boson RG flow is absent [see the first panel of \cref{fig:cft-spec}(a)].

Importantly, the BKT transition should be interpreted as a spin-sector transition rather than a superconducting transition for the following reasons.
First, previous ground-state studies of the $\sigma t$-$J$ model \cite{lu2024sign,zheng2025revealing} demonstrated that the antiferromagnetic spin order is largely decoupled from the doped holes, while the latter form a Fermi-liquid-like state, disfavoring a superconducting interpretation of the finite temperature transition.
Second, we observe that after several RG steps, the partition function tensor eventually loses all elements with odd fermion parity in leg indices. Thus, the fixed-point tensor network is purely bosonic, which is compatible with the interpretation of a spin-sector-only critical point.

\begin{figure}[tb]
    \centering
    \includegraphics[width=0.98\columnwidth]{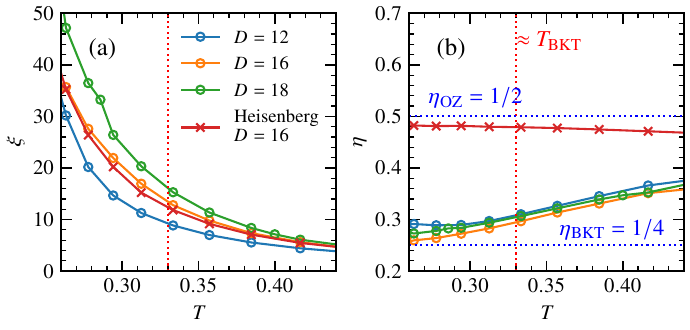}
    \caption{
        (a) Spin correlation length $\xi$ and (b) algebraic exponent $\eta$ [defined in \cref{eq:spincor-fit}] of the $\sigma t$-$J$ model at doping $\delta = 0.12$, for which $T_\text{BKT} \approx 0.33$ estimated from \cref{fig:phases}(a).
        The result is compared with the $\SU(2)$ Heisenberg model ($D = 16$).
    }
    \label{fig:stJ-spincor}
\end{figure}

The most direct evidence comes from the spin correlation. For the $\sigma t$-$J$ model, we examine the in-plane staggered correlation $C^\perp(r) = (-1)^r \braket{B^\perp_{(0,0)(r,0)}}$ [with $i = (0,0)$] by fitting it to the asymptotic form \cite{ding1992phase}
\begin{equation}
    C^\perp(r) = A r^{-\eta} e^{-r/\xi},
    \label{eq:spincor-fit}
\end{equation}
using $2 \le r \le 7$.
\cref{fig:stJ-spincor} shows the fitted algebraic exponent $\eta$ and correlation length $\xi$ at doping $\delta = 0.12$, where the central-charge analysis gives $T_\text{BKT} \approx 0.33$.
Although we do not see the sharp divergence of $\xi$ expected near $T_\text{BKT}$ because the finite iPEPS and CTMRG bond dimensions limit the accessible distances, the fitted exponent $\eta$ clearly decreases toward the BKT value $\eta_\text{BKT} = 1/4$ upon cooling.
In contrast, for the Heisenberg model with full $\SU(2)$ symmetry, the staggered correlation $C(r) = (-1)^r \braket{B_{(0,0)(r,0)}}$ fitted to the same asymptotic form yields an $\eta$ that remains close to the Ornstein-Zernike value $\eta_\text{OZ} = 1/2$ \cite{markivic1991two}.
We have checked that this trend is insensitive to small changes of the largest distance $r$ used for fitting.
Together, these checks support a spin-sector interpretation of the BKT transition in the $\sigma t$-$J$ model.

We note that the comparison to the $t$-$J$ model with a Zeeman field and the $t$-XX model demonstrates that an effective easy-plane spin response alone is not sufficient to explain the robustness against doping of the $\sigma t$-$J$ results.
Only in the $\sigma t$-$J$ model, where the strong hole-spin interference is suppressed, do we find the weak doping dependence of the pseudogap scale $T^*$ and the BKT temperature $T_\text{BKT}$.

\section{PSG-constrained mean-field theory of the \texorpdfstring{$\sigma t$-$J$}{σt-J} model}
\label{sec:mft}

To understand the numerical findings of \cref{sec:num-result}, we develop a slave-fermion mean-field theory of the $\sigma t$-$J$ model based on spin-charge separation \cite{jayaprakash1989mean,yoshioka1989slave,lee2006doping}. The projected electron operator is decomposed as
\begin{equation}
    \tilde{c}_{i\sigma} = b_{i\sigma}\, f_i^\dagger\,,
    \label{eq:parton}
\end{equation}
with $b_{i\sigma}$ a bosonic spinon, $f_i$ a fermionic holon, and $\sigma = \pm 1$ for spin up and down, subject to the local no-double-occupancy constraint $\sum_\sigma b_{i\sigma}^\dagger b_{i\sigma} + f_i^\dagger f_i = 1$.
In terms of the partons, the $\sigma t$-$J$ model can be rewritten exactly within the constrained Hilbert space as
\begin{equation}
    H_{\sigma t\text{-}J}
    = t \sum_{\braket{ij}, \sigma} \sigma\, \hat{\chi}_{ij,\sigma}\, \hat{\kappa}_{ij}^{\dagger}
    + h.c.
    - \frac{J}{2} \sum_{\braket{ij}} \hat{\Delta}_{ij}^{\dagger} \hat{\Delta}_{ij},
    \label{eq:mft-H}
\end{equation}
where $\hat{\kappa}_{ij} = f_i^\dagger f_j$, $\hat{\chi}_{ij,\sigma} = b_{i\sigma}^\dagger b_{j\sigma}$, and $\hat{\Delta}_{ij} = \sum_\sigma \sigma\, b_{i\sigma} b_{j,-\sigma}$ denote the holon-hopping, spinon-hopping, and singlet spinon-pairing bond operators, respectively, obeying $\hat{\kappa}_{ji}=\hat{\kappa}^{\dagger}_{ij}$, $\hat{\chi}_{ji,\sigma}=\hat{\chi}^{\dagger}_{ij,\sigma}$, and $\hat{\Delta}_{ji}=-\hat{\Delta}_{ij}$, with each NN bond counted once in the sums.
The mean-field theory replaces these bond operators by their expectation values, $\kappa_{ij}\equiv\braket{\hat{\kappa}_{ij}}$, $\chi_{ij,\sigma}\equiv\braket{\hat{\chi}_{ij,\sigma}}$, and $\Delta_{ij}\equiv\braket{\hat{\Delta}_{ij}}$, whose lattice pattern defines the ansatz.
Rather than postulating this pattern phenomenologically, in \cref{sec:psg} we show that the PSG classification narrows the symmetric spinon ans\"atze down to $32$ classes, and the correlations observed numerically select just two of them.
In \cref{sec:mf-solution} we then solve the resulting mean-field theory self-consistently and show that it captures the numerical findings of the previous sections: the doping-enhanced in-plane antiferromagnetism, the spin-sector BKT transition, and the pseudogap-like magnetic response.

\subsection{\texorpdfstring{Projective symmetry group analysis and mean-field ansatz}{Projective symmetry group analysis and mean-field ansatz}}
\label{sec:psg}

The $\sigma t$ term breaks the conventional time-reversal symmetry while preserving the $\U(1)$ spin rotation about the $z$ axis, as shown in Appendix~\ref{app:stJ}, which also identifies a composite antiunitary symmetry that does survive. We classify the symmetric spinon mean-field ans\"atze by the PSGs of the square-lattice space group, without imposing time reversal \cite{wen2002quantum,yang2016schwinger}; these PSGs strictly constrain the mean-field amplitudes allowed on each bond.
The projective construction reflects the gauge redundancy of the parton representation. The physical spin operator $\bm{S}_i=\tfrac{1}{2}\sum_{\alpha\beta}b^{\dagger}_{i,\alpha}\bm{\sigma}_{\alpha\beta}b_{i,\beta}$ is invariant under the site-dependent gauge transformation $b_{i,\sigma}\rightarrow b_{i,\sigma}e^{i\phi_i}$, so a physical symmetry $g$ acts on the spinons only up to a gauge transformation $G_g(i)\equiv e^{i\phi_g(i)}\in\U(1)$,
\begin{equation}
    g\,\hat{G}_g (b_{i,\sigma}) = G_g(i)\, b_{g(i),\sigma}\,.
    \label{eq:extended}
\end{equation}
The set of all such $\U(1)$-extended operations that leave a given mean-field ansatz invariant forms its PSG, which refines the classification of spin-liquid states sharing the same physical symmetries \cite{wen2002quantum}. Its subgroup of pure gauge transformations is the invariant gauge group (IGG).

On the square lattice the space group is generated by the translations $T_x$ and $T_y$ and the mirror reflections $\sigma_s$ about the $x$ axis and $\sigma_d$ about the diagonal. When implemented projectively, each defining relation of the space group need hold only up to an element of the IGG. For $\mathrm{IGG}=\mathbb{Z}_2$, the case realized below, the consistency conditions can be solved with an explicit gauge fixing, and as derived in Appendix~\ref{app:psg} all gauge-inequivalent PSGs take the form
\begin{equation}
    \begin{aligned}
        G_x(x,y) & =1\,,                              \\
        G_y(x,y) & =(-1)^{p_{xy}x}\,,                 \\
        G_s(x,y) & =(-1)^{p_x x+p_y y}\, i^{\,p_s}\,, \\
        G_d(x,y) & =(-1)^{p_{xy}xy}\, i^{\,p_d}\,,
    \end{aligned}
    \label{eq:psgsolution}
\end{equation}
where $G_x$, $G_y$, $G_s$, and $G_d$ denote the gauge factors accompanying $T_x$, $T_y$, $\sigma_s$, and $\sigma_d$; $(x,y)$ are the integer site coordinates. Each PSG is therefore characterized by the five $\mathbb{Z}_2$ labels $p_{xy}$, $p_x$, $p_y$, $p_s$, $p_d$, each $=0,1$, yielding $2^{5}=32$ algebraic PSGs. Conventional time reversal is odd on the $\sigma t$ term, as shown in Appendix~\ref{app:stJ}, and is therefore not included among our generators, whereas the general Schwinger-boson classification of Ref.~\cite{yang2016schwinger} includes it.

Each algebraic PSG in turn dictates where the spinon mean fields $\chi_{ij,\sigma}$ and $\Delta_{ij}$ may reside on the lattice and which phases they may carry. For every link there exist space-group elements that map it onto itself or onto its reverse, and demanding invariance of the ansatz under the corresponding $\U(1)$-extended operations either forbids the mean field on that link or pins its phase. Carrying out this analysis for the NN and NNN bonds, as detailed in Appendix~\ref{app:psg}, leads to the constraints summarized in \cref{tab:psgconstraints}.

\begin{table}[tb]
    \caption{PSG constraints on the pairing ($\Delta$) and hopping ($\chi$) mean fields for the NN and NNN bonds of the square lattice. The conditions apply whenever the corresponding mean field is nonzero on that bond; all label equations are understood modulo 2.}
    \label{tab:psgconstraints}
    \begin{ruledtabular}
        \begin{tabular}{lll}
            Bond & Mean field & PSG constraint                           \\
            \colrule
            NN   & $\Delta$   & $p_x+p_y=1$,\quad $p_s=p_x$              \\
            NN   & $\chi$     & $p_x=0$,\quad $\chi^{*}/\chi=(-1)^{p_y}$ \\
            NNN  & $\Delta$   & $p_{xy}=1$,\quad $p_d=1$                 \\
            NNN  & $\chi$     & $p_{xy}=0$,\quad $\chi\in\mathbb{R}$     \\
        \end{tabular}
    \end{ruledtabular}
\end{table}

The reduced spin symmetry of the $\sigma t$-$J$ model enters the analysis in one more way. Since the $\sigma t$ term breaks the $\SU(2)$ spin rotation to $\U(1)$, the hopping amplitudes of the two spin species are no longer tied to each other, enlarging the space of permissible ans\"atze. The two species are nevertheless not independent. On the bipartite square lattice with NN hopping, the $\sigma t$-$J$ model possesses an exact composite $\mathbb{Z}_2$ symmetry $\mathcal{X}$, a global spin flip combined with the staggered sublattice sign $(-1)^{x_i+y_i}$, defined in \cref{eq:Xsym}. This symmetry enforces $\chi_{ij,\uparrow}=-\chi_{ij,\downarrow}$ on the NN bonds and $\chi_{ij,\uparrow}=\chi_{ij,\downarrow}$ on the NNN bonds, both derived in Appendix~\ref{app:stJ}.

Previous tensor-network results for the $\sigma t$-$J$ model found that the NN hopping and pairing correlations, together with the NNN hopping for each spin species, are all nonzero \cite{zheng2025revealing}. These observations restrict the PSG directly through \cref{tab:psgconstraints}. First, the coexistence of the NN pairing and the NN hopping reduces the IGG to $\mathbb{Z}_2$ \cite{read1991large}, because a uniform gauge transformation $b_{i,\sigma}\rightarrow e^{i\phi}b_{i,\sigma}$ leaves every $\chi$ invariant but multiplies every $\Delta$ by $e^{2i\phi}$, so that only $\phi=0,\pi$ survive. Moreover, \cref{tab:psgconstraints} requires $p_x=0$ for the NN hopping, upon which the NN pairing fixes $p_y=1$ and $p_s=p_x=0$. Second, the nonzero NNN hopping requires $p_{xy}=0$, which excludes all $\pi$-flux states. Only the label $p_d$ then remains free. With the labeling $\mathbb{Z}_2(p_{xy}\pi,\,p_d\pi)$, the $\sigma t$-$J$ model is restricted to the two zero-flux classes
\begin{equation}
    \begin{aligned}
        \mathbb{Z}_2(0,0)   & :\ (p_{xy},p_x,p_y,p_s,p_d)=(0,0,1,0,0)\,, \\
        \mathbb{Z}_2(0,\pi) & :\ (p_{xy},p_x,p_y,p_s,p_d)=(0,0,1,0,1)\,,
    \end{aligned}
    \label{eq:labels0}
\end{equation}
both subject to the mean-field constraints
\begin{equation}
    \text{NN } \chi_{ij,\sigma}\in i\mathbb{R}\,,\quad
    \text{NN } \Delta_{ij}\in i\mathbb{R}\,,\quad
    \text{NNN } \Delta_{ij}=0\,.
    \label{eq:constraints00}
\end{equation}
Here the purely imaginary NN hopping follows from \cref{tab:psgconstraints} with $p_y=1$, whereas the overall phase of the NN pairing is fixed by the residual global $\U(1)$ transformation, which leaves the gauge-fixed $G_g(i)$ of \cref{eq:psgsolution} invariant while rotating the phase of $\Delta$. We use this freedom to align the pairing phase with the imaginary axis.
In this way, the numerical data select the relevant classes out of the many symmetry-allowed possibilities. This eases a practical limitation of parton mean-field constructions, namely that the PSG classification by itself typically leaves too many candidate classes to identify the effective mean-field theory of a given microscopic model.

The bond pattern of the ansatz follows explicitly from the PSG.
For the $\mathbb{Z}_2(0,0)$ class, substituting the labels of \cref{eq:labels0} into \cref{eq:psgsolution} gives
\begin{equation}
    \begin{aligned}
        G_x(x,y) & =1\,,        &
        G_y(x,y) & =1\,,
        \\
        G_s(x,y) & =(-1)^{y}\,, &
        G_d(x,y) & =1\,.
    \end{aligned}
    \label{eq:psg00}
\end{equation}
The PSG determines the relative phases of the mean fields on all bonds once they are fixed on a single bond. Up to the real amplitudes determined self-consistently below, we take the representative values on the bond from the origin to $(1,0)$,
\begin{equation}
    \chi_{(1,0),\uparrow}=i\,,
    \quad
    \Delta_{(1,0)}=i\,,
    \label{eq:fiducial}
\end{equation}
consistent with the constraint \cref{eq:constraints00}. For $g=\sigma_d$, which maps the $y$ bond onto the $x$ bond, and for $g=T_x$, the invariance condition \cref{eq:invariance} together with \cref{eq:chitrans,eq:deltatrans} gives
\begin{equation}
    \begin{aligned}
        \chi_{(0,1),\uparrow}       & =G_d^{*}(0,0)\, \chi_{(1,0),\uparrow}\, G_d(1,0)=i\,,  \\
        \chi_{(-1,0)(0,0),\uparrow} & =G_x^{*}(-1,0)\, \chi_{(1,0),\uparrow}\, G_x(0,0)=i\,, \\
        \Delta_{(0,1)}              & =G_d(0,0)\, \Delta_{(1,0)}\, G_d(1,0)=i\,,             \\
        \Delta_{(-1,0)(0,0)}        & =G_x(-1,0)\, \Delta_{(1,0)}\, G_x(0,0)=i\,.
    \end{aligned}
    \label{eq:bonds00}
\end{equation}
Since the translational gauge factors are trivial, the same pattern repeats on every site. Both spinon mean fields are uniform and take the same purely imaginary value on all $x$ and $y$ bonds, oriented along the positive lattice directions.

The $\mathbb{Z}_2(0,0)$ PSG thus dictates the mean-field ansatz
\begin{equation}
    \begin{aligned}
        \braket{\hat{\kappa}_{i, i + \hat{x}/\hat{y}}}       & = i \kappa,
        \\
        \braket{\hat{\Delta}_{i, i + \hat{x}/\hat{y}}}       & = i \Delta,
        \\
        \braket{\hat{\chi}_{i, i + \hat{x}/\hat{y}, \sigma}} & = i \sigma \chi,
    \end{aligned}
    \label{eq:mft-ansatz}
\end{equation}
with real $\kappa$, $\chi$, and $\Delta$. The $\mathbb{Z}_2(0,\pi)$ ansatz differs only in the $d$-wave-like sign structure of its pairing, $\braket{\hat{\Delta}_{i,i+\hat{y}}}=-\braket{\hat{\Delta}_{i,i+\hat{x}}}$, as constructed in Appendix~\ref{app:mf0pi}.
Notably, the spinons carry both the resonating-valence-bond (RVB) pairing $\Delta$ and the hopping $\chi$; the latter is generated by the $\sigma t$ term and takes opposite signs for the two spin species, reflecting the reduced spin symmetry. As discussed above, this opposite-sign structure is not an additional assumption but is enforced by the exact composite symmetry $\mathcal{X}$.
At this stage the PSG and correlation constraints have left two candidate zero-flux states. Comparing their mean-field energies, we find that the $\mathbb{Z}_2(0,0)$ state has the distinctly lower energy. Moreover, upon spinon condensation the $\mathbb{Z}_2(0,\pi)$ state would develop a uniform magnetization along the $z$ axis on top of the in-plane antiferromagnetic order, which is not observed in the iPEPS ground state \cite{zheng2025revealing}. Both criteria, established in Appendix~\ref{app:mf0pi}, consistently select the $\mathbb{Z}_2(0,0)$ state, and we adopt it in the following.
We emphasize that the PSG constraints are formulated for the zero-temperature mean-field ansatz. However, through its self-consistency equations the mean-field theory built on this ansatz extends directly to finite temperature and, as shown in \cref{sec:mf-solution}, captures the physics found numerically.

\subsection{\texorpdfstring{Self-consistent solution and finite-temperature physics}{Self-consistent solution and finite-temperature physics}}
\label{sec:mf-solution}

Decoupling \cref{eq:mft-H} with the ansatz \cref{eq:mft-ansatz} and Fourier transforming with $b_{i,\sigma}=N^{-1/2}\sum_{k}b_{k\sigma}e^{-ik\cdot r_i}$ (and similarly for $f_i$), where $N$ is the number of lattice sites, yields the mean-field Hamiltonian $H^{\mathrm{MF}}_{\sigma t\text{-}J}=H_f+H_b$. The holon part reads
\begin{equation}
    H_f=-4t\chi\sum_{k}\mathcal{E}_k\, f^{\dagger}_{k}f_{k}
    -\mu\sum_{k}\left(f^{\dagger}_{k}f_{k}-\delta\right),
    \label{eq:Hf}
\end{equation}
with the lattice function $\mathcal{E}_k=\sin k_x+\sin k_y$, while the spinon part, written in terms of the Nambu spinor $\Psi_k=\big(b_{k\uparrow},\,b^{\dagger}_{-k\downarrow}\big)^{\mathsf{T}}$, reads
\begin{equation}
    \begin{aligned}
        H_b={}           & \sum_{k}\Psi^{\dagger}_{k}\,\mathcal{H}_k\,\Psi_{k}
        +J\Delta^{2}N-8tN\chi\kappa-(2-\delta)\lambda N\,,                                       \\
        \mathcal{H}_k={} & \begin{pmatrix}
                               \lambda-2t\kappa\,\mathcal{E}_k & J\Delta\,\mathcal{E}_k          \\
                               J\Delta\,\mathcal{E}_k          & \lambda-2t\kappa\,\mathcal{E}_k
                           \end{pmatrix},
    \end{aligned}
    \label{eq:Hb}
\end{equation}
where the chemical potentials $\mu$ and $\lambda$ enforce the average holon and spinon densities $\delta$ and $1-\delta$.
The two spin species carry opposite kinetic terms $-2t\kappa\sigma\,\mathcal{E}_k$, which merge into the equal diagonal entries of $\mathcal{H}_k$ upon relabeling $k\rightarrow-k$ in the spin-$\downarrow$ sector, since $\mathcal{E}_{-k}=-\mathcal{E}_k$. The holon part is already diagonal, and a bosonic Bogoliubov transformation of $\mathcal{H}_k$ yields the holon and Bogoliubov-spinon dispersions
\begin{equation}
    \begin{aligned}
        E^{f}_{k} & = -4t\chi\, \mathcal{E}_k - \mu\,,                                                                 \\
        E^{b}_{k} & = \sqrt{\left(\lambda - 2t\kappa\, \mathcal{E}_k\right)^{2} - J^{2}\Delta^{2}\mathcal{E}_k^{2}}\,.
    \end{aligned}
    \label{eq:mft-dispersion}
\end{equation}

Explicitly, the Bogoliubov quasiparticles read
\begin{equation}
    \gamma_{k\uparrow}=u_k\, b_{k\uparrow}+v_k\, b^{\dagger}_{-k\downarrow},
    \quad
    \gamma_{k\downarrow}=u_k\, b_{-k\downarrow}+v_k\, b^{\dagger}_{k\uparrow}\,,
    \label{eq:bogoliubov}
\end{equation}
with the coefficients
\begin{equation}
    u_k=\sqrt{\frac{\varepsilon_k+E^{b}_{k}}{2E^{b}_{k}}},
    \quad
    v_k=\sgn\left(\Delta_k\right)\sqrt{\frac{\varepsilon_k-E^{b}_{k}}{2E^{b}_{k}}},
    \label{eq:ukvk}
\end{equation}
where $\varepsilon_k\equiv\lambda-2t\kappa\,\mathcal{E}_k$ and $\Delta_k\equiv J\Delta\,\mathcal{E}_k$ are the diagonal and off-diagonal entries of $\mathcal{H}_k$, so that $u_k^{2}-v_k^{2}=1$ and $u_kv_k=\Delta_k/\big(2E^{b}_{k}\big)$.
The free energy then follows from the standard fermionic and bosonic mode sums,
\begin{equation}
    \begin{aligned}
        F={} & -\frac{1}{\beta}\sum_{k}\ln\left(1+e^{-\beta E^{f}_{k}}\right)
        +\frac{2}{\beta}\sum_{k}\ln\left(1-e^{-\beta E^{b}_{k}}\right)        \\
             & +\sum_{k}E^{b}_{k}
        +J\Delta^{2}N-(2-\delta)\lambda N-8tN\chi\kappa+\mu\delta N\,,
    \end{aligned}
    \label{eq:freeenergy}
\end{equation}
where $\beta=1/T$ is the inverse temperature. The saddle-point equations follow from the stationarity of $F$ with respect to the five parameters, $\partial F/\partial\lambda=\partial F/\partial\mu=\partial F/\partial\kappa=\partial F/\partial\chi=\partial F/\partial\Delta=0$, which yield, respectively,
\begin{subequations}
    \begin{align}
         & \frac{1}{N}\sum_{k}\frac{\lambda-2t\kappa\,\mathcal{E}_k}{E^{b}_{k}}
        \coth\left(\frac{\beta E^{b}_{k}}{2}\right)=2-\delta\,,
        \label{eq:saddlelambda}                                                                           \\
         & \frac{1}{N}\sum_{k}\frac{1}{1+e^{\beta E^{f}_{k}}}=\delta\,,
        \label{eq:saddlemu}                                                                               \\
         & \frac{1}{N}\sum_{k}\frac{\mathcal{E}_k\left(\lambda-2t\kappa\,\mathcal{E}_k\right)}{E^{b}_{k}}
        \coth\left(\frac{\beta E^{b}_{k}}{2}\right)
        +4\chi=0\,,
        \label{eq:saddlekappa}                                                                            \\
         & \frac{1}{N}\sum_{k}\frac{\mathcal{E}_k}{1+e^{\beta E^{f}_{k}}}+2\kappa=0\,,
        \label{eq:saddlechi}                                                                              \\
         & 2-\frac{1}{N}\sum_{k}\frac{J\,\mathcal{E}_k^{2}}{E^{b}_{k}}
        \coth\left(\frac{\beta E^{b}_{k}}{2}\right)=0\,.
        \label{eq:saddledelta}
    \end{align}
\end{subequations}

\begin{figure}[tb]
    \centering
    \includegraphics[width=\columnwidth]{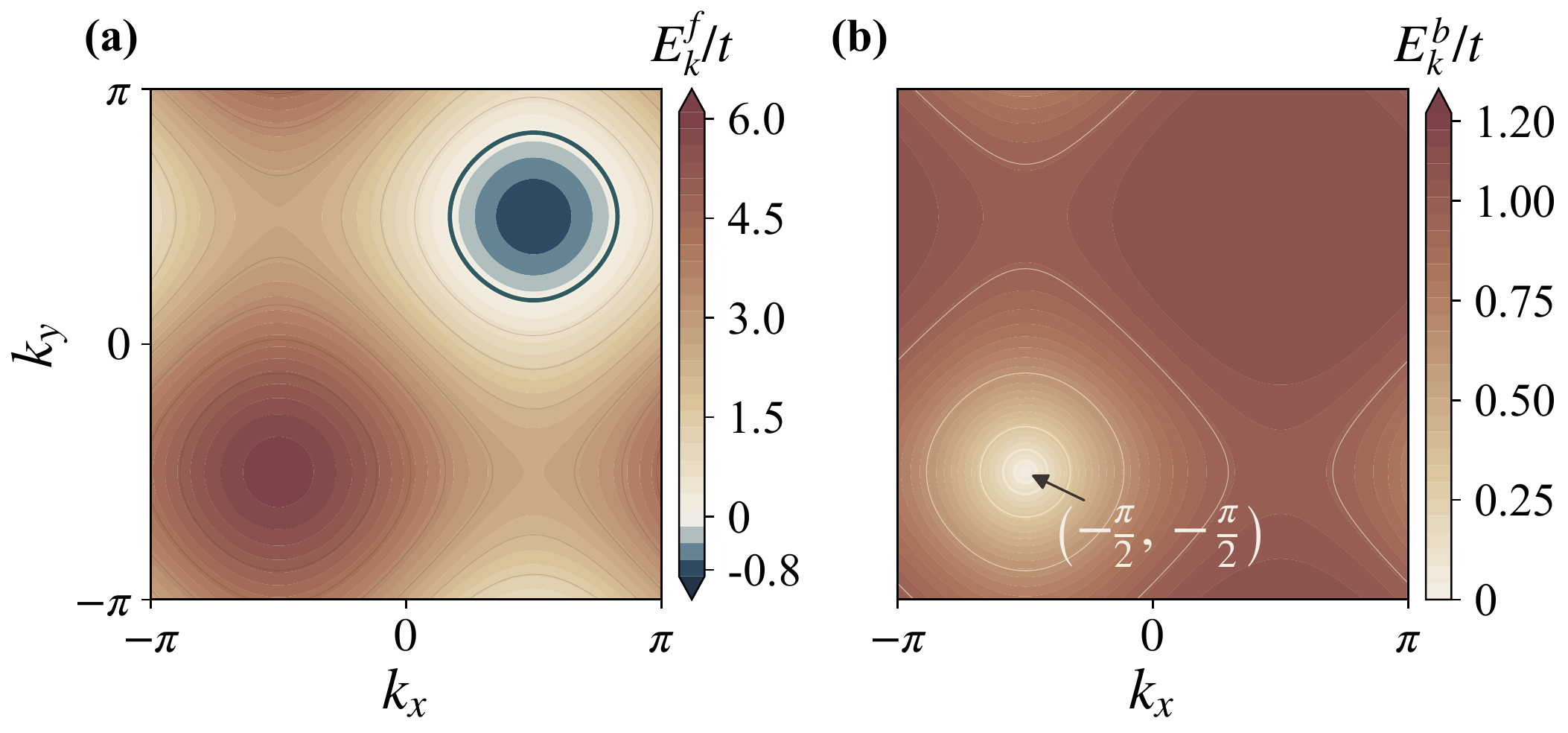}
    \caption{Zero-temperature mean-field dispersions of the $\mathbb{Z}_2(0,0)$ state at $\delta=1/12$ and $t/J=2$, obtained from the self-consistent solution of \cref{eq:saddlelambda,eq:saddlemu,eq:saddlekappa,eq:saddlechi,eq:saddledelta}. (a)~Holon dispersion $E^{f}_{k}/t$. The dark contour marks the Fermi surface, which encloses a pocket centered at $(\pi/2,\pi/2)$ of area $\delta\times A_{\mathrm{BZ}}$. (b)~Bogoliubov-spinon dispersion $E^{b}_{k}/t$. The spectrum is minimal at $Q=\left(-\pi/2,-\pi/2\right)$ (arrow) and at the symmetry-related point $-Q$, where the gap closes at $T=0$ and the spinons undergo Bose-Einstein condensation.}
    \label{fig:dispersions}
\end{figure}

Solving \cref{eq:saddlelambda,eq:saddlemu,eq:saddlekappa,eq:saddlechi,eq:saddledelta} at $T=0$ for the representative parameters $\delta=1/12$ and $t/J=2$ (we set $t=1$ as the energy unit for the mean-field results), we obtain the holon and spinon dispersions shown in \cref{fig:dispersions}. The holon dispersion $E^{f}_{k}$ exhibits a Fermi pocket of area $\delta\times A_{\mathrm{BZ}}$ centered at $(\pi/2,\pi/2)$, the maximum of $\mathcal{E}_k$, where $A_{\mathrm{BZ}}$ is the Brillouin-zone area, as shown in \cref{fig:dispersions}(a). The Bogoliubov-spinon dispersion $E^{b}_{k}$ has its low-lying modes at the momenta $\pm Q$ with $Q\equiv(-\pi/2,-\pi/2)$, where $|\mathcal{E}_k|$ is maximal, as shown in \cref{fig:dispersions}(b). Although $E^{b}_{k}$ is even in $k$, the two spin species carry opposite dispersions $-2t\kappa\sigma\,\mathcal{E}_k$ in \cref{eq:Hb}; since the self-consistent solution has $\kappa<0$, the low-lying spectral weight of the $\uparrow$-spinons ($\downarrow$-spinons) resides at $Q$ ($-Q$). The solution further shows that the spinons undergo Bose-Einstein condensation at $T=0$, in close analogy with the conventional Schwinger-boson theory of the Heisenberg model \cite{arovas1988functional,auerbach1988spin}. The difference is that the $\uparrow$- and $\downarrow$-spinons condense at the two distinct momenta $Q$ and $-Q$, respectively.

For the physical electrons, the parton decomposition \cref{eq:parton} factorizes the electron Green's function $G^{c}_{\sigma}(\tau;r_i-r_j)=-\braket{T_{\tau}\,\tilde{c}_{i\sigma}(\tau)\,\tilde{c}^{\dagger}_{j\sigma}(0)}$ at the mean-field level into a spinon propagator accompanied by a backward-propagating holon. The condensation of the $\sigma$-spinon at momentum $\sigma Q$ replaces the spinon propagator by its condensate weight and rigidly relabels the holon momentum, so that in frequency-momentum space
\begin{equation}
    G^{c}_{\sigma}(\omega,k)\propto -\braket{f^{\dagger}_{k+\sigma Q}(\omega)\, f_{k+\sigma Q}(\omega)}.
    \label{eq:electronGk}
\end{equation}
Up to this momentum shift the electron and holon Green's functions are proportional, indicating a Fermi-liquid phase. The spin-$\uparrow$ and spin-$\downarrow$ electron pockets have the same volume $\delta\times A_{\mathrm{BZ}}$ as the holon pocket of \cref{fig:dispersions}(a), but are shifted from it by $-\sigma Q$ and thus located at distinct momenta.

We now analyze the magnetic order that develops when the spinons condense, together with its fate at finite temperature, which is governed by the topology of the order-parameter manifold. At the condensation momenta $\pm Q$, where $\mathcal{E}_{\pm Q}=\mp 2$, the spinon part \cref{eq:Hb} decomposes into two decoupled Nambu sectors. The first, $\left(b_{Q\uparrow},\, b^{\dagger}_{-Q\downarrow}\right)$, is governed by
\begin{equation}
    H_b\big|_{\mathrm{soft}}=
    \begin{pmatrix}
        b^{\dagger}_{Q\uparrow} & b_{-Q\downarrow}
    \end{pmatrix}
    \begin{pmatrix}
        \lambda+4t\kappa & -2J\Delta        \\
        -2J\Delta        & \lambda+4t\kappa
    \end{pmatrix}
    \begin{pmatrix}
        b_{Q\uparrow} \\ b^{\dagger}_{-Q\downarrow}
    \end{pmatrix},
    \label{eq:Hbsoft}
\end{equation}
The second, $\left(b_{Q\downarrow},\, b^{\dagger}_{-Q\uparrow}\right)$, is governed by the same matrix with $\lambda+4t\kappa$ replaced by $\lambda-4t\kappa$. Since $\kappa<0$, the first sector lies below the second by $8t|\kappa|$. We retain only this soft sector and project out the gapped one, and this projection is what determines the order-parameter manifold below. Diagonalizing \cref{eq:Hbsoft} gives a gapless condensed eigenmode, the symmetric combination of $b_{Q\uparrow}$ and $b^{\dagger}_{-Q\downarrow}$. The two condensate amplitudes are therefore locked to a single complex number $z$, $\braket{b_{Q\uparrow}}=\sqrt{N}\,z$ and $\braket{b_{-Q\downarrow}}=\sqrt{N}\,z^{*}$. The condensed spinon density is then $n_0=\big(|\braket{b_{Q\uparrow}}|^{2}+|\braket{b_{-Q\downarrow}}|^{2}\big)/N=2|z|^{2}$. Since $\braket{b_{-Q\uparrow}}=\braket{b_{Q\downarrow}}=0$ in the projected theory, the real-space condensate amplitudes are
\begin{equation}
    \braket{b_{i\uparrow}}=z\, e^{-iQ\cdot r_i}\,,
    \quad
    \braket{b_{i\downarrow}}=z^{*} e^{iQ\cdot r_i}\,,
    \label{eq:condspinor}
\end{equation}
and the resulting local magnetization reads
\begin{equation}
    \braket{S^{z}_i}=0\,,
    \quad
    \braket{S^{+}_i}=\left(z^{*}\right)^{2}e^{2iQ\cdot r_i}\propto(-1)^{x_i+y_i}\,,
    \label{eq:spinorder}
\end{equation}
where $e^{2iQ\cdot r_i}=(-1)^{x_i+y_i}$. The condensation produces antiferromagnetic order lying in the $xy$ plane, with no ordered moment along the $z$ axis and with staggered moment $m_s=|\braket{S^{+}_i}|=|z|^{2}=n_0/2$.
The zero-temperature state is therefore a Fermi liquid coexisting with easy-plane antiferromagnetic long-range order.

\begin{figure}[tb]
    \centering
    \includegraphics[width=0.75\columnwidth]{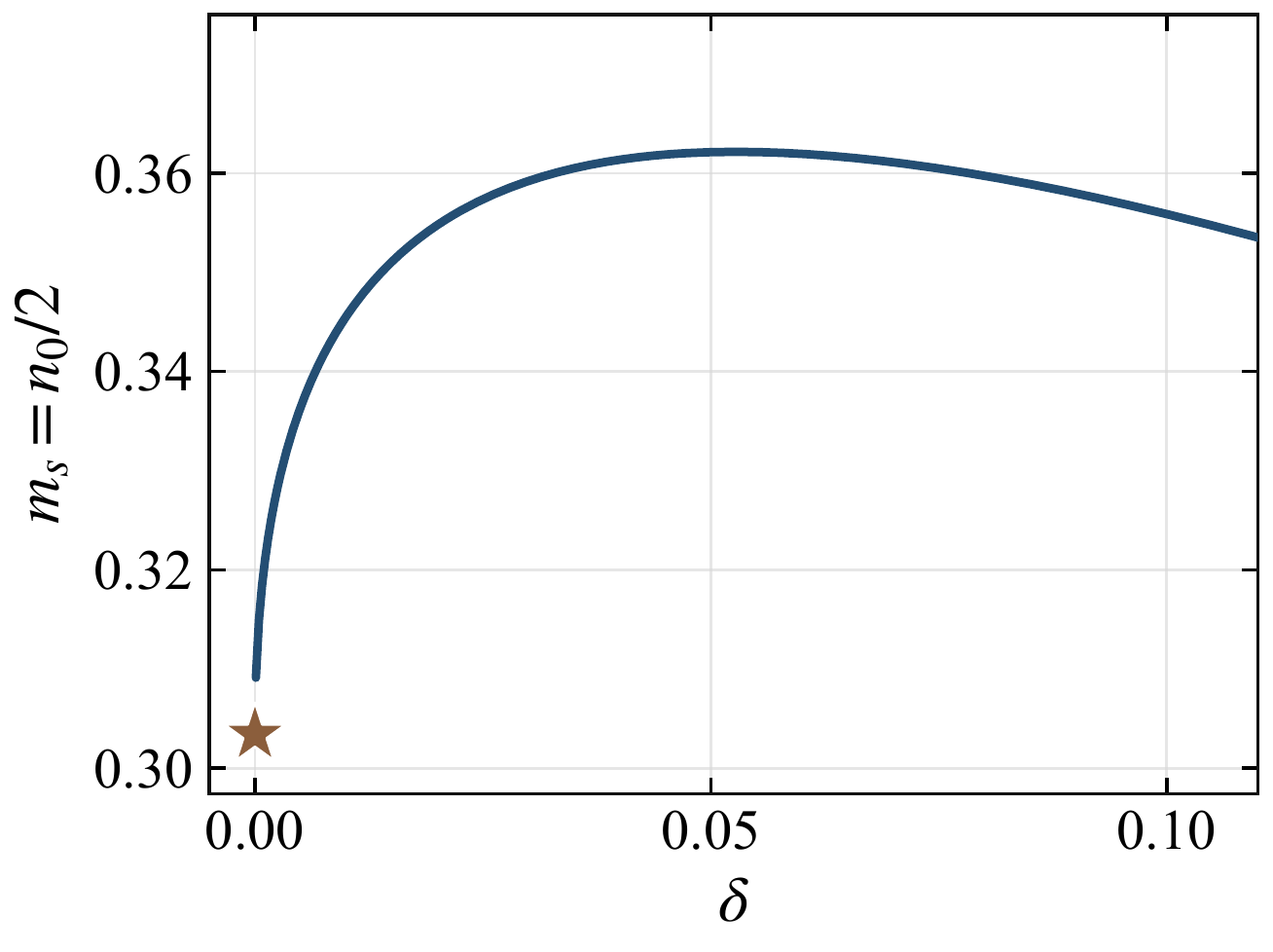}
    \caption{
        Zero-temperature staggered in-plane moment $m_s = n_0/2$ of the $\mathbb{Z}_2(0,0)$ mean-field state as a function of doping $\delta$, obtained from the self-consistent solution of \cref{eq:saddlelambda,eq:saddlemu,eq:saddlekappa,eq:saddlechi,eq:saddledelta} at $t/J = 2$.
        The star at $\delta = 0$ marks the Schwinger-boson mean-field result $m_s \simeq 0.303$ of the half-filled Heisenberg antiferromagnet \cite{arovas1988functional,sarker1989bosonic}.
        Dilute doping enhances the in-plane antiferromagnetic moment relative to the undoped limit.
    }
    \label{fig:mft-ms}
\end{figure}

The doping dependence of the staggered moment is shown in \cref{fig:mft-ms}.
As $\delta \to 0$ the moment smoothly approaches the half-filled Schwinger-boson benchmark $m_s \simeq 0.303$ \cite{arovas1988functional,sarker1989bosonic}, marked by the star, and then \emph{increases} upon doping, reaching its maximum around $\delta \approx 0.05$ before decreasing slowly.
Through the $\sigma t$ term, the doped holes generate a spin-weighted spinon hopping that is maximized by antiparallel spins lying in the $xy$ plane, so the hole kinetic energy cooperates with the superexchange in stabilizing the in-plane antiferromagnetism instead of frustrating it, as the semiclassical analysis in Appendix~\ref{app:mf00} shows. This is where the local hole-spin interplay of the $\sigma t$-$J$ model differs qualitatively from that of the $t$-$J$ model.
This mean-field behavior agrees well with the doping-enhanced $xy$-plane spin correlations found in \cref{fig:stJ-tJ-polaron-doping}.

The topology of the spinon condensate further explains the finite-temperature BKT-like behavior suggested in \cref{sec:stj-bkt}.
For any nonzero doping only the single sector $\left(b_{Q\uparrow},\, b^{\dagger}_{-Q\downarrow}\right)$ condenses, so the order parameter \cref{eq:spinorder} lives on the circle $S^{1}\cong\U(1)$ parametrized by the phase of $z^{2}$, corresponding to the residual spin rotation about the $z$ axis left intact by the $\sigma t$ term. The doped $\sigma t$-$J$ model is therefore expected to follow the standard BKT phenomenology of a two-dimensional $\U(1)$ order parameter \cite{mermin1966absence,berezinskii1971destruction,berezinskii1972destruction,kosterlitz1973ordering}, with quasi-long-range in-plane antiferromagnetic order below $T_\text{BKT}$ and true long-range order only at $T=0$.
In the undoped limit, by contrast, the holon bond amplitude vanishes together with the hole density, as shown in \cref{sec:n0}, so the gap splitting $8t|\kappa|$ closes and \emph{both} Nambu sectors condense, the second with its own locked amplitude, $\braket{b_{Q\downarrow}}=\sqrt{N}\, v$ and $\braket{b_{-Q\uparrow}}=-\sqrt{N}\, v^{*}$. The condensate amplitudes generalize \cref{eq:condspinor} to
\begin{equation}
    \begin{aligned}
        \braket{b_{i\uparrow}}
         & = z\, e^{-iQ\cdot r_i}-v^{*}e^{iQ\cdot r_i}\,,
        \\
        \braket{b_{i\downarrow}}
         & = v\, e^{-iQ\cdot r_i}+z^{*}e^{iQ\cdot r_i}\,,
    \end{aligned}
    \label{eq:undopedspinor}
\end{equation}
and the local magnetization becomes purely staggered,
\begin{equation}
    \begin{aligned}
        \braket{S^{+}_i} & =\left[\left(z^{*}\right)^{2}-v^{2}\right](-1)^{x_i+y_i}\,, \\
        \braket{S^{z}_i} & =-2\operatorname{Re}\left(zv\right)(-1)^{x_i+y_i}\,,
    \end{aligned}
    \label{eq:undopedorder}
\end{equation}
with magnitude $|\bm{m}_s|=|z|^{2}+|v|^{2}=n_0/2$ independent of the direction. At fixed $n_0$ the condensate $(z,v)$ spans the three-sphere $S^{3}$. Moreover, since the spinon ansatz now contains only pairing bonds, the IGG is enlarged from $\mathbb{Z}_2$ to a staggered $\U(1)$, with opposite phases on the two sublattices, which acts on the condensate as $z\rightarrow z\cos\phi-iv^{*}\sin\phi$ and $v\rightarrow v\cos\phi+iz^{*}\sin\phi$ while leaving the magnetization \cref{eq:undopedorder} invariant. The physical order-parameter manifold at $\delta=0$ is therefore $S^{3}/\U(1)\cong S^{2}$, the full N\'eel sphere of the $\SU(2)$-symmetric Heisenberg limit. Since $S^{2}$ admits no vortices, the undoped system has no finite-temperature transition \cite{mermin1966absence,chakravarty1989two}. Doping thus creates the order-parameter topology needed for a finite-temperature BKT scenario by collapsing the manifold from $S^{2}$ down to $\U(1)$ through the sector projection.
This contrast matches the loop-TNR phase diagram in \cref{fig:phases}, where the extended spin-sector critical regime at finite doping terminates as $\delta \to 0$.

\begin{figure}[tb]
    \centering
    \includegraphics[width=0.82\columnwidth]{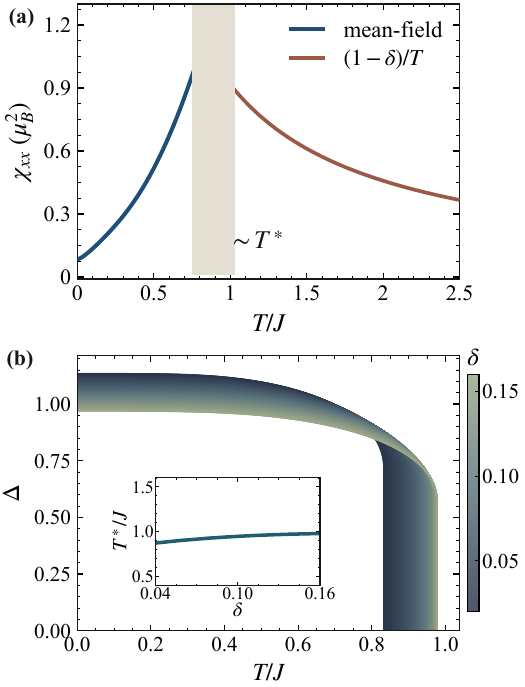}
    \caption{
        Mean-field magnetic response of the $\sigma t$-$J$ model.
        (a) In-plane spin susceptibility $\chi_{xx}$ at $\delta = 1/12$ (blue), in units of $\mu_B^2$, compared with the Curie's law behavior $(1-\delta)/T$ of independent local moments (red), valid above $T^*$.
        The shaded region marks the crossover around $T^*$, whose detailed structure is beyond the mean-field description.
        (b) RVB pairing amplitude $\Delta$ as a function of temperature for various dopings $\delta$ (color bar); $\Delta$ vanishes at $T^*$.
        Inset: $T^*/J$ versus $\delta$, showing only a weak doping dependence, since $T^*$ is controlled by the superexchange $J$.
    }
    \label{fig:mft-chi}
\end{figure}

Finally, the same mean-field theory accounts for the pseudogap-like magnetic response identified in \cref{sec:pseudogap}.
The relevant mean-field scale is set by the RVB pairing amplitude. When the self-consistency equations are solved at finite temperature, $\Delta(T)$ decreases upon heating and vanishes at a scale $T^*$, which we identify with the onset of short-range RVB pairing, as shown in \cref{fig:mft-chi}(b).

To compute the corresponding magnetic response, we couple the spinons to a uniform transverse Zeeman field, $-\mu_B g B_x\sum_i S^{x}_i$, with the convention $\mu_B=1$ and $g=2$ throughout the mean-field results, so that magnetization and susceptibility are measured in units of $\mu_B$ and $\mu_B^{2}$.
This adds to the spinon Hamiltonian \cref{eq:Hb} the term $-B_x\sum_{k}\big(b^{\dagger}_{k\uparrow}b_{k\downarrow}+b^{\dagger}_{k\downarrow}b_{k\uparrow}\big)$, which mixes the two spin species and does not commute with the singlet pairing.
The Bogoliubov diagonalization of the resulting quadratic Hamiltonian couples the modes at $k$ and $-k$, yielding for each such pair the four quasiparticle poles $\pm E_{+,k}$ and $\pm E_{-,k}$ with
\begin{equation}
    \begin{aligned}
        E_{\pm,k} & \equiv\sqrt{4t^{2}\kappa^{2}\mathcal{E}_k^{2}-\Delta_k^{2}+\lambda^{2}+B_x^{2}\pm 2 M_k}\,,               \\
        M_k       & \equiv\sqrt{4t^{2}\kappa^{2}\mathcal{E}_k^{2}\lambda^{2}+B_x^{2}\left(\lambda^{2}-\Delta_k^{2}\right)}\,.
    \end{aligned}
    \label{eq:Epm}
\end{equation}
The free energy of the spinon sector then reads
\begin{equation}
    F_x=\frac{2}{\beta}\sum_{k>0}\sum_{m=\pm}\ln\left(1-e^{-\beta E_{m,k}}\right)
    +\sum_{k>0}\sum_{m=\pm}E_{m,k}+C\,,
    \label{eq:Fx}
\end{equation}
where the momentum sum is restricted to half of the Brillouin zone so that each pair $(k,-k)$ is counted once, and $C$ collects the field-independent constants of \cref{eq:Hf,eq:Hb}. Differentiation with respect to the field gives the $x$ magnetization
\begin{equation}
    \begin{aligned}
        M_x & =-\frac{\partial F_x}{\partial B_x} \\
            & =B_x\sum_{k>0}\sum_{m=\pm}
        \frac{\coth\left(\frac{1}{2}\beta E_{m,k}\right)}{E_{m,k}}\left(m R_k-1\right),
    \end{aligned}
    \label{eq:Mx}
\end{equation}
with $R_k\equiv\left(\Delta_k^{2}-\lambda^{2}\right)/M_k$. In the limit $B_x\rightarrow 0$ of $\chi_{xx}=M_x/B_x$, the mean-field in-plane susceptibility takes the closed form
\begin{equation}
    \chi_{xx}=\sum_{k}
    \frac{\coth\left(\frac{1}{2}\beta E^{b}_{k}\right)}{E^{b}_{k}}
    \left(\frac{\lambda^{2}-\Delta_k^{2}}{2t\kappa\lambda\,\mathcal{E}_k}-1\right),
    \label{eq:chixx}
\end{equation}
where on the lines $\mathcal{E}_{k}=0$ the summand is understood as its limiting value $\beta\,\mathrm{csch}^{2}\!\left(\beta\lambda/2\right)$. On these lines the two branches $E_{\pm,k}$ become degenerate, and only the sum of their individually divergent ($\propto B_{x}^{-1}$) contributions to \cref{eq:Mx} remains finite.
Below $T^*$ the progressive formation of spin singlets suppresses the tendency toward uniform spin polarization, and correspondingly $\chi_{xx}$ decreases upon cooling, as shown by the blue curve in \cref{fig:mft-chi}(a).
Above $T^*$, the RVB pairs break apart and the spin background reduces to a density $1 - \delta$ of essentially independent local moments, which lose their short-range singlet correlations and give rise to the Curie's law behavior $\chi_{xx} \simeq (1-\delta)/T$, shown by the red curve in \cref{fig:mft-chi}(a).
The resulting maximum of $\chi_{xx}$ around $T^*$ thus reproduces the pseudogap-like magnetic crossover of \cref{fig:mag-chi}. The detailed shape of the crossover connecting the two regimes is beyond the mean-field description, and is indicated by the shaded region in \cref{fig:mft-chi}(a).
Moreover, since $T^*$ is the temperature at which the local RVB pairing dissolves, it is governed by the superexchange $J$ and depends only weakly on doping, as shown in the inset of \cref{fig:mft-chi}(b). This agrees with the weak doping dependence of the pseudogap onset temperature found numerically in \cref{fig:mag-chi}.

\section{Conclusion and outlook}
\label{sec:conclusion}

We investigated the $\sigma t$-$J$ model at finite temperature using iPEPS algorithms, and found that it shares the spin-pseudogap thermodynamics of the original $t$-$J$ model: the magnetic susceptibility has a maximum at $T^*$ and the specific heat develops a broad peak.
The PSG-constrained slave-fermion mean-field theory reproduces this spin-pseudogap behavior, with $T^*$ set by the dissolution of the short-range RVB pairing.

Alongside these similarities, the two models differ sharply in how robust the finite-temperature signatures are against doping.
In the $\sigma t$-$J$ model, doped holes preserve and even enhance the $xy$ component of the antiferromagnetic correlations, the pseudogap scale $T^*$ changes only weakly over the studied doping range, and the specific-heat peak remains robust in the intermediate-doping regime.
Loop-TNR further reveals a stable $c \approx 1$ regime with compactified-boson spectra over a broad finite-doping range, and the accompanying spin-correlation analysis supports its interpretation as a spin-sector BKT transition at $T_\text{BKT}<T^*$.
By contrast, the original $t$-$J$ model does not show a stable BKT flow in the same parameter regime, while the $t$-XX model and the Zeeman-field $t$-$J$ model show that reducing the spin symmetry can make spin-sector BKT behavior visible but does not by itself make the finite-temperature scales robust against doping.
Taken together, these results indicate that the robust pseudogap behavior and broad spin BKT regime of the $\sigma t$-$J$ model rely not only on an effective easy-plane spin response, but also on suppressing the disturbance to the spin background caused by hole-spin interplay.
Consistently, the same contrast appears already in the ground state, where the stripe and superconducting orders of the $t$-$J$ and $t$-$t^{\prime}$-$J$ models are absent in the $\sigma t$-$J$ model and doping instead yields a Fermi-liquid-like state of small hole pockets coexisting with the antiferromagnetic spin background \cite{zheng2025revealing,lu2024sign}.

The much simpler behavior of the $\sigma t$-$J$ model found in this work, compared with the $t$-$J$ model, has a transparent microscopic origin \cite{lu2024sign,zheng2025revealing}.
For the original $t$-$J$ model, the partition function has the exact form \cite{wu2008sign}
\begin{equation}
    Z_{t\text{-}J} = \tr \rho_{t\text{-}J} = \sum_C \tau_C W[C],
    \label{eq:tj-partition-func}
\end{equation}
where $C$ sums over all spin and hole configurations in which holes move in closed loops, and $W[C] \ge 0$. The Berry phase $\tau_C = \pm 1$ consists of two contributions,
\begin{equation}
    \tau_{C} = (-1)^{N_{\text{ex}}^{h}[C]}
    (-1)^{N_{h}^{\downarrow}[C]},
    \label{eq:phase-string}
\end{equation}
where $N_{\text{ex}}^{h}[C]$ and $N_{h}^{\downarrow}[C]$ count the number of exchanges with holes or down spins as a hole moves on the closed loop $C$, respectively.
In other words, besides the conventional Fermi statistical signs $N_{\text{ex}}^{h}[C]$ between holes, a moving hole also picks up a sign $+1$ or $-1$ each time it exchanges with an up or a down spin.
This accumulated, irreparable string of signs $N_{h}^{\downarrow}[C]$, known as the \emph{phase string}, constitutes a singular source of many-body quantum interference beyond the conventional Fermi statistics of the holes \cite{sheng1996phase,weng1997phase,wu2008sign}.
The spin background thereby imposes a strong sign frustration on the hole kinetic energy, and relieving this frustration forces the holes to reorganize the surrounding spins locally.
This is why a small density of holes already renormalizes the antiferromagnet severely, destroys the antiferromagnetic long-range order at low doping, and helps spawn the complex landscape of competing orders in the $t$-$J$ model.
On the other hand, the spin-dependent sign of the $\sigma t$ hopping in \cref{eq:stJ-hamiltonian} exactly compensates the phase string signs: the partition function of the $\sigma t$-$J$ model still has the form of \cref{eq:tj-partition-func}, but $\tau_C$ reduces to just $(-1)^{N_{\text{ex}}^{h}[C]}$ \cite{lu2024sign}.
Freed from this kinetic frustration by construction, the hole motion in the $\sigma t$-$J$ model remains compatible with the $xy$-plane antiferromagnetic order. This is ultimately why the antiferromagnetic correlations, the pseudogap $T^*$, and the spin-sector BKT criticality all remain robust against doping.

Finally, we discuss open directions for future investigation.
Increasing bond dimensions in iPEPS, CTMRG and loop-TNR would further refine the phase boundary and conformal spectra, especially near the edge of the BKT region.
Algorithms that directly produce a single-layer iPEPO form of $e^{-\beta H}$, such as the cluster expansion method \cite{vanhecke2021symmetric, vanhecke2023simulating, demeyer2026lowering}, should help reduce computational cost and improve the quality of tensor-network renormalization inputs.
Such improvements will also enable more reliable fits of the correlation length $\xi$ and exponent $\eta$ over longer distances, making it possible to resolve the expected BKT divergence of $\xi$ as $T$ approaches $T_\text{BKT}$.
It would also be useful to extend the calculations to lower temperatures and wider doping ranges to search for possible charge ordering, pairing, or other instabilities beyond the spin-sector criticality suggested by the present data.

Two further open directions concern the implications of the fractionalized framework of \cref{sec:mft}.
The first is the electron spectral function of the $\sigma t$-$J$ model.
In this framework, the physical electron is a composite mode of a bosonic spinon and a fermionic holon rather than an elementary quasiparticle \cite{zhang2023crossover}.
In the ground state at finite doping the spinons are condensed. The electron propagator therefore inherits the coherence of the gapless holons, consistent with the coherent single-particle propagation and the Fermi-liquid-like behavior found at zero temperature \cite{lu2024sign,zheng2025revealing}.
Above $T_\text{BKT}$, however, the spinon condensate loses its phase coherence, and the electron, given by the convolution of the spinon and holon propagators, need not survive as a sharp excitation. Its low-energy spectral weight is then expected to be suppressed, which would appear as a spectroscopic pseudogap analogous to the one observed by ARPES in the cuprates \cite{ding1996spectroscopic,loeser1996excitation}.
The second is a possible pairing instability beyond the mean-field level. Gauge fluctuations may induce a pair-density-wave (PDW) fluctuation, similar to the mechanism discussed for the $t$-$J$ model on the honeycomb lattice \cite{miao2025spin}, in which case PDW fluctuations would be a natural consequence of spin-charge separation.
Signatures of fluctuating PDW order have indeed been reported in the square-lattice $t$-$J$ model \cite{zheng2025competing,yue2024pseudogap} and in the $t$-$t^{\prime}$ Hubbard model at finite temperature \cite{li2026fluctuating}.
As tensor network methods continue to advance, computing the finite-temperature electron spectral function of the $\sigma t$-$J$ model and searching for the PDW fluctuation are natural targets for future work.

\begin{acknowledgments}
    ZYY thanks Sander De Meyer and Yintai Zhang for discussions on the NTU algorithm for finite temperature, and Chenqi Meng and Atsushi Ueda for discussions on loop-TNR for fermionic systems.
    This work is supported by funding from Hong Kong’s Research Grants Council (CRF C7015-24G, CRS HKU701/24). JXZ was supported by the European Research Council (ERC) under the European Union’s Horizon 2020 research and innovation program (Grant Agreement No.\ 853116, acronym TRANSPORT). JXZ was also supported in part by grant NSF PHY-2309135 to the Kavli Institute for Theoretical Physics (KITP).
\end{acknowledgments}

\appendix

\section{Spin BKT transition in pure spin XXZ and Heisenberg models}
\label{app:xxz}

\begin{figure}[tb]
    \centering
    \includegraphics[width=\columnwidth]{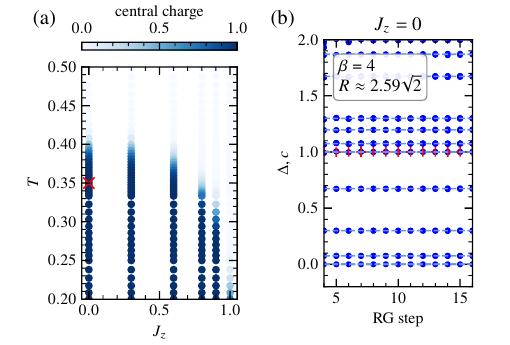}
    \caption{
        (a) Finite-temperature central charge diagram (after RG step 10) for the XXZ model ($D = 12$) when $0 \le J_z < 1$. The red cross shows $T_\text{BKT} \approx 0.33$ \cite{ding1992phase,harada1997universal} at $J_z = 0$.
        (b) CFT spectrum for $J_z = 0$ and $\beta = 4$, which matches the compactified boson CFT with fitted radius $R \approx 2.59 \sqrt{2}$.
    }
    \label{fig:xxz-kt}
\end{figure}

\begin{figure}[tb]
    \centering
    \includegraphics[width=0.9\columnwidth]{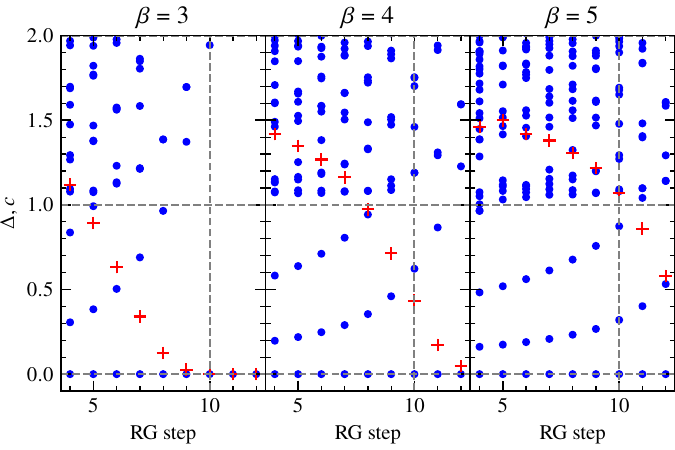}
    \caption{
        Transient RG flow of the square lattice Heisenberg model with $J = J_z = 1$, obtained with iPEPS bond dimensions $D = 16$. $\SU(2)$ symmetry is imposed on the tensors.
    }
    \label{fig:heis-kt}
\end{figure}

\begin{figure}[tb]
    \centering
    \includegraphics[width=0.9\columnwidth]{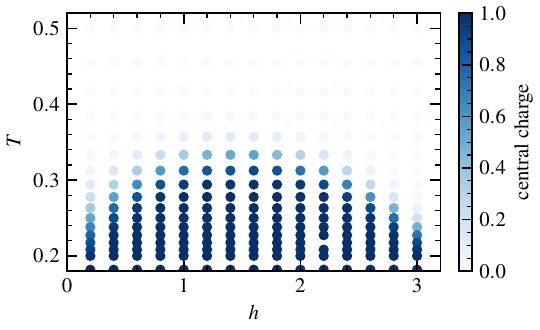}
    \caption{
        Finite-temperature central charge diagram for the Heisenberg model ($D = 12$) with a uniform magnetic field $h$ along the $z$ axis.
    }
    \label{fig:heis-kt-mag}
\end{figure}

In this appendix, we benchmark the loop-TNR BKT diagnostics on pure spin models.
We start from the spin-1/2 XXZ Hamiltonian
\begin{equation}
    H = \sum_{\braket{ij}} \left[
        J (S^x_i S^x_j + S^y_i S^y_j) + J_z S^z_i S^z_j
        \right].
    \label{eq:xxz-hamiltonian}
\end{equation}
Below we set $J = 1$. When $0 \le J_z < 1$, the exchange interaction has easy-plane anisotropy and the XXZ model is known to undergo a BKT transition at finite temperature $T_\text{BKT} > 0$ \cite{ding1992phase,biagini1996quantum,ueda2026global}.
\cref{fig:xxz-kt} shows the thermal phase diagram obtained from the central charge extracted by loop-TNR.
The transition temperature appears to be highest when $J_z = 0$, and gradually decreases to 0 as $J_z$ approaches 1.
In the low-temperature critical region, the scaling dimension spectrum can be fitted by the compactified-boson CFT in \cref{eq:boson-cft} with an appropriate radius $R$, as illustrated for $J_z = 0$ in \cref{fig:xxz-kt}(b).

The isotropic point $J_z = 1$ (Heisenberg model) provides a useful contrast.
Here the full $\SU(2)$ spin rotation symmetry is restored, and no finite-temperature phase transition is expected due to the Mermin-Wagner theorem \cite{mermin1966absence}.
Loop-TNR nevertheless produces a transient unstable RG flow, shown in \cref{fig:heis-kt}. The central charge starts around $c \approx 1.5$ and gradually drifts away from the initial spectrum instead of forming a stable compact-boson plateau.
This behavior is similar to the classical two-dimensional Heisenberg model on the square lattice \cite{ueda2022tensor}, and is inherited by the $t$-$J$ model (without Zeeman field) at very small doping [\cref{fig:phases}(b) and \cref{fig:cft-spec}(b)].

Finally, adding a uniform magnetic field $H_h = -h \sum_i S^z_i$ to the Heisenberg model also explicitly reduces the spin rotation symmetry to $\U(1)$ and makes the transverse spin correlation effectively easy-plane.
As shown in \cref{fig:heis-kt-mag}, as $h$ increases from 0, the Heisenberg model develops a finite-temperature BKT region (with $c = 1$ and compactified boson CFT spectrum) at numerically accessible temperatures.
However, if $h$ becomes too large, the spins will polarize in the $z$ direction, reducing the AFM correlation in the $xy$ plane. Thus the BKT transition temperature falls again, as observed when $h \gtrsim 2$ in \cref{fig:heis-kt-mag}.

\section{Details of the slave-fermion mean field theory for \texorpdfstring{$\sigma t$-$J$}{σt-J}}
\label{app:mft}

\subsection{Projective symmetry group analysis}
\label{app:psg}

As introduced in \cref{sec:psg} of the main text, owing to the $\U(1)$ gauge redundancy of the Schwinger-boson representation, a symmetry $g$ in the group of \emph{physical} symmetries $\mathrm{SG}$ acts on the bosonic spinons only projectively. It may be accompanied by a gauge transformation $G_g(i)\equiv e^{i\phi_g(i)}\in \U(1)$ according to \cref{eq:extended}.
Distinct spin liquids may share the same physical symmetries and thus cannot be told apart by broken symmetries or local order parameters; they are instead distinguished by the gauge transformations $G_g(i)$ that must accompany each symmetry operation to leave their mean-field ans\"atze invariant. The set of all $\U(1)$-extended operations under which a given ansatz $u_{i,j}$, defined on the lattice links, is invariant,
\begin{equation}
    g\,\hat{G}_g (u_{i,j}) = u_{i,j} \ ,
    \label{eq:invariance}
\end{equation}
forms the projective symmetry group (PSG), which provides a refined classification of symmetric spin-liquid phases \cite{wen2002quantum,wang2006spin,yang2016schwinger}. A special subgroup of the PSG is the invariant gauge group (IGG), consisting of all \emph{gauge} transformations $\hat{\mathcal{G}}$ that by themselves leave the ansatz invariant,
\begin{equation}
    \mathrm{IGG}=\left\{\hat{\mathcal{G}} \,\middle|\, \hat{\mathcal{G}}\left(u_{i,j}\right)=u_{i,j}\right\}.
    \label{eq:iggdef}
\end{equation}
The gauge transformations $G_g(i)$ are defined only up to a gauge rotation of the ansatz and up to multiplication by IGG elements; PSGs related in this way describe the same physical state. The classification below therefore enumerates the gauge-inequivalent solutions after an explicit gauge fixing.

\subsubsection{Schwinger-boson PSGs on the square lattice}
\label{sec:classification}

Consider a square lattice whose sites are labeled by pairs of integer coordinates $(x,y)$. The space group is generated by the translations $T_x$ and $T_y$, the mirror reflection $\sigma_s$ about the $x$ axis, and the mirror reflection $\sigma_d$ about the diagonal,
\begin{equation}
    \begin{aligned}
        T_x      & :(x,y)\rightarrow(x+1,y)\,, &
        T_y      & :(x,y)\rightarrow(x,y+1)\,,
        \\
        \sigma_s & :(x,y)\rightarrow(x,-y)\,,  &
        \sigma_d & :(x,y)\rightarrow(y,x)\,,
    \end{aligned}
    \label{eq:generators}
\end{equation}
which satisfy the defining relations
\begin{equation}
    \begin{gathered}
        T_x^{-1}T_y^{-1}T_xT_y=1\,,\\
        \sigma_s T_x=T_x\sigma_s\,,\qquad
        \sigma_s T_y=T_y^{-1}\sigma_s\,,\\
        \sigma_d T_x=T_y\sigma_d\,,\qquad
        \sigma_d T_y=T_x\sigma_d\,,\\
        \sigma_s^2=1\,,\qquad
        \sigma_d^2=1\,,\qquad
        \left(\sigma_s\sigma_d\right)^4=1\,.
    \end{gathered}
    \label{eq:relations}
\end{equation}
When implemented projectively according to \cref{eq:extended}, each defining relation requires the product of the associated gauge factors to equal an element of the IGG rather than the identity. For $\mathrm{IGG}=\mathbb{Z}_2$, the case of interest below, every relation thus introduces a $\mathbb{Z}_2$ sign, and solving the resulting algebraic equations site by site determines the admissible sets $\{G_g(i)\}$ \cite{wen2002quantum,yang2016schwinger}. Adopting the gauge fixing
\begin{equation}
    G_x(x,y)=1\,,\qquad G_y(0,y)=1\,,
    \label{eq:gaugefixing}
\end{equation}
where $G_x$, $G_y$, $G_s$, and $G_d$ denote the gauge factors accompanying $T_x$, $T_y$, $\sigma_s$, and $\sigma_d$, one obtains all the gauge-inequivalent PSGs in the form quoted in \cref{eq:psgsolution} of the main text, characterized by the five $\mathbb{Z}_2$ labels $p_{xy}$, $p_x$, $p_y$, $p_s$, $p_d=0,1$; as noted there, since time reversal is not imposed as a generator, the $64$ classes of Ref.~\cite{yang2016schwinger} reduce to $2^5=32$.

\subsubsection{PSG constraints on the mean-field ans\"atze}
\label{sec:constraints}

Not every algebraic PSG admits a mean-field realization on the lattice. The presence of certain mean-field parameters imposes constraints on the PSG and, conversely, a given PSG dictates how the mean-field parameters may be chosen. Here we consider the spinon pairing and hopping mean fields,
\begin{equation}
    \Delta_{0,i} = \braket{\epsilon_{\alpha\beta}\, b_{0,\alpha} b_{i,\beta}},
    \quad
    \chi_{0,i,\sigma} = \braket{b_{0,\sigma}^{\dagger} b_{i,\sigma}},
    \label{eq:mfdef}
\end{equation}
where $\epsilon_{\alpha\beta}$ is the antisymmetric tensor with $\epsilon_{\uparrow\downarrow}=1$ and the pair of sites $(0,i)$ specifies the lattice link. The pairing mean field $\Delta$ corresponds to local resonating-valence-bond (RVB) singlet pairing, while the hopping mean field $\chi$ represents local spinon hopping for each spin species. Under the PSG operation induced by $g$ they transform as
\begin{align}
    \chi_{0,i,\sigma}\rightarrow g\circ \chi_{0,i,\sigma}
     & =G_g^{*}(0)\, G_g(i)\, \chi_{0,g(i),\sigma}\,,
    \label{eq:chitrans}                               \\
    \Delta_{0,i}\rightarrow g\circ \Delta_{0,i}
     & =G_g(0)\, G_g(i)\, \Delta_{0,g(i)}\,,
    \label{eq:deltatrans}
\end{align}
the hopping field being of the particle-hole and the pairing field of the particle-particle type. This difference is precisely what reduces the IGG to $\mathbb{Z}_2$ when both channels are present. A uniform transformation $b_{i,\sigma}\rightarrow e^{i\phi}b_{i,\sigma}$ leaves every $\chi$ invariant but multiplies every $\Delta$ by $e^{2i\phi}$, so that only $\phi=0,\pi$ survive.

The constraints then follow from a simple observation. For every link there exist space-group elements that map the link onto itself or onto its reverse. Combining \cref{eq:chitrans,eq:deltatrans} with the antisymmetry $\Delta_{i,j}=-\Delta_{j,i}$ of the singlet pair and the Hermiticity $\chi_{i,j,\sigma}=\chi^{*}_{j,i,\sigma}$, such operations impose self-consistency conditions that either forbid the mean field on that link or restrict its phase. For a generic link from the origin to $i=(x,y)$, the combined operation $T_x^{x}T_y^{y}(\sigma_s\sigma_d)^2$ reverses the link, and the PSG solution \cref{eq:psgsolution} yields
\begin{equation}
    \begin{aligned}
        -\Delta_{0,i} & =W_i\,\Delta_{0,i}\,,
        \quad
        \chi^{*}_{0,i,\sigma}=W_i\,\chi_{0,i,\sigma}\,,    \\
        W_i           & =(-1)^{p_{xy}xy+(p_x+p_y)(x+y)}\,,
    \end{aligned}
    \label{eq:reversal}
\end{equation}
so that $\Delta_{0,i}$ can exist only if $W_i=-1$, while a nonzero $\chi_{0,i,\sigma}$ must satisfy $\chi^{*}/\chi=W_i$. Links lying on the axes or on the diagonals satisfy further conditions of the same type, obtained from the additional operations $T_y^{y}\sigma_s$ and $\sigma_d\sigma_s\sigma_d$ (for an axial link) and $\sigma_s T_x^{x}T_y^{-x}\sigma_d\sigma_s$ and $\sigma_d$ (for a diagonal link). Working out these conditions for the nearest-neighbor link, e.g., $i=(1,0)$, and for the next-nearest-neighbor link, e.g., $i=(1,1)$, one obtains the constraints summarized in \cref{tab:psgconstraints} of the main text.

\subsubsection{Application to the \texorpdfstring{$\sigma t$-$J$}{σt-J} model}
\label{app:stJ}

The $\sigma t$-$J$ model is defined by the spin-dependent hopping term in \cref{eq:stJ-hamiltonian} together with the exchange term in \cref{eq:tJ-hamiltonian}. We use $\sigma=\pm 1$ for spin up and down. As in \cref{sec:mft}, the electron operator is decomposed according to \cref{eq:parton}, with $b_{i\sigma}$ the bosonic spinon and $f_i$ the fermionic holon. The no-double-occupancy constraint $\sum_\sigma c^\dagger_{i\sigma} c_{i\sigma} \le 1$ translates to $\sum_{\sigma}b^{\dagger}_{i\sigma}b_{i\sigma}+f^{\dagger}_i f_i = 1$ \cite{jayaprakash1989mean,yoshioka1989slave,lee2006doping}.

We first point out that the $\sigma t$-hopping term explicitly breaks both the conventional time-reversal symmetry and the $\SU(2)$ spin-rotation symmetry.
To show that the time reversal symmetry is lost, it is convenient to introduce the electron spinor $\mathbf{c}_i=(c_{i\uparrow},c_{i\downarrow})^{\mathsf{T}}$, so that
\begin{align}
    H_t          & = -t\sum_{\braket{ij}}
    \tilde{\mathbf{c}}^{\dagger}_{i}
    \tilde{\mathbf{c}}_{j} + h.c. \ ,
    \\
    H_{\sigma t} & = -t\sum_{\braket{ij}}
    \tilde{\mathbf{c}}^{\dagger}_{i} \sigma^z
    \tilde{\mathbf{c}}_{j} + h.c. \ ,
    \\
    \mathbf{S}_i & = \frac{1}{2} \mathbf{c}^\dagger_i
    \boldsymbol{\sigma} \mathbf{c}_i \ ,
\end{align}
where $\tilde{\mathbf{c}}$ is $\mathbf{c}$ restricted to the no-double-occupancy subspace. Explicitly, time reversal is the antiunitary operation $\Theta$ that complex conjugates all c-numbers and reverses the spins, acting on the electron operators as
\begin{equation}
    \Theta \mathbf{c}_i \Theta^{-1}
    = i\sigma^{y} \, \mathbf{c}_i.
    \label{eq:time-reversal-action}
\end{equation}
One can verify that $\mathbf{S}_i$ is mapped to $-\mathbf{S}_i$, since
\begin{equation}
    [(i\sigma^y)^\dagger \sigma^a (i\sigma^y)]^* = -\sigma^a
    \quad (a = x, y, z).
\end{equation}
Therefore, both $H_t$ and $H_J$ are invariant under $\Theta$, so the original $t$-$J$ model is time-reversal symmetric.
In contrast, since $(\sigma^{y}\sigma^{z}\sigma^{y})^* = -\sigma^{z}$, we obtain
\begin{equation}
    \Theta H_{\sigma t} \Theta^{-1} = -H_{\sigma t}.
    \label{eq:stJ-time-reversal}
\end{equation}
Therefore, $H_{\sigma t}$ is odd under time reversal, so $\Theta$ alone is not a symmetry of the $\sigma t$-$J$ model.

The loss of $\SU(2)$ spin rotation symmetry is evident from $H_{\sigma t}$ not being invariant under the spin flip $\uparrow \ \leftrightarrow \ \downarrow$, leaving only the $\U(1)$ spin rotation about the $z$ axis intact.
This enlarges the space of permissible mean-field ans\"atze, since the hopping mean fields of the two spin species are no longer tied to each other.
The two spin species are nevertheless not independent. On the bipartite square lattice with NN-only hopping, the $\sigma t$-$J$ model possesses an exact composite $\mathbb{Z}_2$ symmetry that combines a global spin flip with a staggered sublattice sign. We define the operation
\begin{equation}
    \mathcal{X}:\ \tilde{\mathbf{c}}_i
    \mapsto \zeta_i\,\sigma^{x}\,\tilde{\mathbf{c}}_i\,,
    \quad
    \zeta_i=(-1)^{x_i+y_i}\,.
    \label{eq:Xsym}
\end{equation}

The spin flip reverses the sign of the $\sigma t$ term, $\sigma^{x}\sigma^{z}\sigma^{x}=-\sigma^{z}$, which is exactly compensated by $\zeta_i\zeta_j=-1$ on every NN bond, so $H_{\sigma t}$ is invariant; $H_J$ in \cref{eq:tJ-hamiltonian} is invariant as well, since $\mathbf{S}_i\cdot\mathbf{S}_j$ and $n_in_j$ are unchanged under a global spin rotation while the on-site signs $\zeta_i$ cancel, and the no-double-occupancy constraint is manifestly preserved. In the parton representation \cref{eq:parton}, $\mathcal{X}$ acts as $b_{i,\sigma}\rightarrow\zeta_i\,b_{i,-\sigma}$ with $f_i$ untouched, under which the spinon-hopping bond operator of \cref{eq:mft-H} transforms on the NN bonds as
\begin{equation}
    \hat{\chi}_{ij,\sigma}
    \mapsto \zeta_i\zeta_j\,\hat{\chi}_{ij,-\sigma}
    =-\hat{\chi}_{ij,-\sigma}\,,
\end{equation}
while $\hat{\Delta}_{ij}$ and $\hat{\kappa}_{ij}$ are invariant. Demanding that the mean-field ansatz preserve $\mathcal{X}$ (with a trivial accompanying gauge transformation) therefore fixes the relative sign of the two hopping species on the NN bonds,
\begin{equation}
    \chi_{ij,\uparrow}=-\chi_{ij,\downarrow}\,,
    \label{eq:chiflip}
\end{equation}
while leaving $\Delta$ and $\kappa$ unconstrained. (On the NNN bonds $\zeta_i\zeta_j=+1$, so the same argument enforces $\chi_{ij,\uparrow} = \chi_{ij,\downarrow}$ on NNN bonds.)

The staggered sign $\zeta_i$ also yields an antiunitary symmetry.
Acting with $U_{\zeta}:\tilde{\mathbf{c}}_i\mapsto\zeta_i\,\tilde{\mathbf{c}}_i$ alone flips the sign of every NN bond and hence of $H_{\sigma t}$, exactly compensating the sign in \cref{eq:stJ-time-reversal}. The combination $\Theta^{\prime}=U_{\zeta}\,\Theta$ is an exact symmetry of the $\sigma t$-$J$ model.
We do not impose $\Theta^{\prime}$ as an additional generator in the PSG classification of Appendix~\ref{app:psg}, which is carried out with the space group alone.
Instead, we verify it a posteriori on the ans\"atze selected in \cref{sec:psg}. With the staggered sign implemented on the holons, $\Theta^{\prime}$ reverses only the sign of the pairing amplitude in \cref{eq:mft-ansatz}, which is undone by the uniform gauge transformation $b_{i\sigma}\rightarrow i\,b_{i\sigma}$, $f_{i}\rightarrow i\,f_{i}$, and the same holds for the $\mathbb{Z}_2(0,\pi)$ ansatz.
Both candidate states are therefore compatible with $\Theta^{\prime}$, so leaving it out of the classification does not affect the analysis in this work.

As discussed in \cref{sec:psg} of the main text, the numerically observed coexistence of the NN pairing, the NN hopping, and the NNN hopping \cite{zheng2025revealing} reduces the IGG to $\mathbb{Z}_2$ \cite{read1991large}, as explained below \cref{eq:deltatrans}, and restricts the PSG through \cref{tab:psgconstraints} to the two zero-flux classes $\mathbb{Z}_2(0,0)$ and $\mathbb{Z}_2(0,\pi)$ quoted in \cref{eq:labels0}, subject to the bond constraints \cref{eq:constraints00}. The mean-field theory of the $\mathbb{Z}_2(0,0)$ state is constructed and solved in \cref{sec:mf-solution} of the main text; Appendix~\ref{app:mf0pi} constructs the competing $\mathbb{Z}_2(0,\pi)$ state and shows that $\mathbb{Z}_2(0,0)$ has the lower mean-field energy.

It is instructive to relate our gauge to that of Ref.~\cite{yang2016schwinger} by Yang and Wang, which classifies the same states using the generator set $\{T_x,T_y,\sigma_s,C_4\}$, with $C_4=\sigma_d\sigma_s$, and the gauge transformation $G^\text{YW}_g$ is applied after the action of the physical symmetry $g$,
\begin{equation*}
    b_{i,\sigma}
    \to G^\text{YW}_g g (b_{i,\sigma})
    = e^{i\phi_g[g(i)]}b_{g(i),\sigma}.
\end{equation*}
Rewriting their solutions [Eq.~(11) therein, with labels $(p_1,p_2,p_3,p_4,p_7)$] in the convention of \cref{eq:extended} and composing $\sigma_d=C_4\sigma_s$, their gauge factor for the diagonal mirror reads $G^{\text{YW}}_d(x,y)=(-1)^{p_1xy}(-1)^{p_3(x+y)}\,i^{\,p_4+p_7}$, which differs from $G_d$ of \cref{eq:psgsolution} by the staggered factor $(-1)^{p_3(x+y)}$ (the states of interest have $p_y=p_3=1$). The two gauges are connected by the explicit gauge transformation
\begin{equation}
    b_{(x,y),\sigma} \to e^{i\pi/4}\,(-1)^{y}\, b_{(x,y),\sigma}\,,
    \label{eq:gaugemapYW}
\end{equation}
i.e., $\chi_{ij}\rightarrow(-1)^{y_i+y_j}\chi_{ij}$ and $\Delta_{ij}\rightarrow i\,(-1)^{y_i+y_j}\Delta_{ij}$. The staggered factor $(-1)^{y}$ converts their $C_4$ gauge into our $\sigma_d$ gauge, at the harmless cost of an IGG sign $G_y\rightarrow-G_y$, while the global phase $e^{i\pi/4}$ is the residual $\U(1)$ rotation mapping their time-reversal-pinned \emph{real} NN pairing onto our purely imaginary convention in \cref{eq:constraints00}. Under \cref{eq:gaugemapYW} the classification labels correspond as $(p_{xy},p_x,p_y,p_s,p_d)=(p_1,p_2,p_3,p_4,p_4+p_7 \bmod 2)$.

\subsection{\texorpdfstring{Further analysis of the $\mathbb{Z}_2(0,0)$ state}{Further analysis of the Z2(0,0) state}}
\label{app:mf00}

\subsubsection{Semiclassical analysis}
\label{sec:semiclassical}

The easy-plane order can also be understood semiclassically, by evaluating on spin coherent states the two bond amplitudes that acquire mean values in \cref{eq:mft-ansatz}. The pairing $\Delta=\braket{b_{i\uparrow}b_{j\downarrow}-b_{i\downarrow}b_{j\uparrow}}$ and the spin-weighted hopping $\chi=\braket{b^{\dagger}_{i\uparrow}b_{j\uparrow}-b^{\dagger}_{i\downarrow}b_{j\downarrow}}$, the latter being the combination $\sum_{\sigma}\sigma\hat{\chi}_{ij,\sigma}$ generated by the $\sigma t$ term. Let the spin at site $j$ point in the direction $(\theta,\phi)$, i.e.,
\begin{equation}
    b_{j}=\big(e^{-i\phi}\cos(\theta/2),\,\sin(\theta/2)\big)^{\mathsf{T}}
\end{equation}
If the spin at site $i$ is in $+z$ direction, $b_{i}=(1,\,0)^{\mathsf{T}}$, one finds $\Delta=\sin(\theta/2)$ and $\chi=e^{-i\phi}\cos(\theta/2)$. The pairing is maximized by antiparallel and the hopping by parallel alignment, so along the $z$ axis the two channels compete.
If the spin at $i$ is in $+x$ direction, $b_{i}=\tfrac{1}{\sqrt{2}}(1,\,1)^{\mathsf{T}}$, one finds instead
\begin{equation}
    \begin{aligned}
        \Delta & =\frac{1}{\sqrt{2}}\sin(\theta/2)-\frac{1}{\sqrt{2}}e^{-i\phi}\cos(\theta/2)\,, \\
        \chi   & =\frac{1}{\sqrt{2}}e^{-i\phi}\cos(\theta/2)-\frac{1}{\sqrt{2}}\sin(\theta/2)\,,
    \end{aligned}
    \label{eq:cohx}
\end{equation}
which gives $(|\Delta|,|\chi|)=(0,0)$ for parallel in-plane spins [$(\theta,\phi)=(\pi/2,0)$] but $(1,1)$ for antiparallel in-plane spins [$(\theta,\phi)=(\pi/2,\pi)$]. In the $xy$ plane both channels are simultaneously maximized by antiparallel spins and cooperate rather than compete. The in-plane antiferromagnetic correlation is therefore the strongest when both fields are present, which is the semiclassical, hopping-driven origin of the easy-plane order in \cref{eq:spinorder}; in the ordinary $t$-$J$ model, the spin-independent hopping instead favors locally parallel spins and competes with the superexchange. The same picture distinguishes the NN and NNN bonds. In the in-plane collinear antiferromagnetic configuration, a NN bond $\ket{\rightarrow}_i\ket{\leftarrow}_j$ carries the maximal spin-weighted hopping $\braket{b^{\dagger}_{i\uparrow}b_{j\uparrow}-b^{\dagger}_{i\downarrow}b_{j\downarrow}}=-1$, while a NNN bond $\ket{\rightarrow}_i\ket{\rightarrow}_j$ (same sublattice) carries the maximal ordinary hopping $\braket{b^{\dagger}_{i\uparrow}b_{j\uparrow}+b^{\dagger}_{i\downarrow}b_{j\downarrow}}=1$, consistent with the numerically observed mean-field structure invoked in Appendix~\ref{app:stJ}.

\subsubsection{Doping dependence of the condensate density}
\label{sec:n0}

Since the staggered moment is tied to the condensate through $m_s=n_0/2$, as given by \cref{eq:spinorder}, the doping dependence of $n_0$ measures the strength of the zero-temperature magnetic order. We now compute $n_0(\delta)$ analytically in the thermodynamic limit at $T=0$, by reducing the saddle-point equations \eqref{eq:saddlelambda}--\eqref{eq:saddledelta} to a single scalar equation. Because all integrands depend on $k$ only through $x\equiv\mathcal{E}_k$, the momentum sums reduce to one-dimensional integrals, $\frac{1}{N}\sum_{k} f(\mathcal{E}_k)\rightarrow\int_{-2}^{2}dx\,\rho(x)f(x)$, over the density of states $\rho(x)=\pi^{-2}K(1-x^{2}/4)$, the convolution of the two arcsine distributions of $\sin k_x$ and $\sin k_y$, with $K$ the complete elliptic integral of the first kind.

At $T=0$ the holon occupation is a step function. Since $E^{f}(x)=-4t\chi x-\mu$ decreases with $x$ (for $\chi>0$), the occupied states satisfy $x>\xi\equiv-\mu/(4t\chi)$, and the two holon equations \eqref{eq:saddlemu} and \eqref{eq:saddlechi} become
\begin{equation}
    \delta=\int_{\xi_\delta}^{2}dx\,\rho(x)\,,
    \quad
    \kappa_\delta=-\frac{1}{2}\int_{\xi_\delta}^{2}dx\, x\,\rho(x)\,,
    \label{eq:holon}
\end{equation}
which fix the Fermi-surface cutoff $\xi_\delta$ and the holon bond amplitude $\kappa_\delta$ independently of the spinon sector; in particular $\kappa_\delta<0$ throughout (at low doping $\kappa_\delta\simeq-\delta+\tfrac{\pi}{2}\delta^{2}$), the sign that selects the soft Nambu sector in \cref{sec:mf-solution}.

In the spinon sector, the gap at $\pm Q$ is $E^{b}_{Q}=\sqrt{(\lambda+4t\kappa)^{2}-(2J\Delta)^{2}}$, and condensation, $E^{b}_{Q}=0$, requires (on the stable branch for $\Delta>0$)
\begin{equation}
    \lambda+4t\kappa=2J\Delta\,.
    \label{eq:condcondition}
\end{equation}
The condensate contributes the classical energy density $f_0=n_0\left(\lambda+4t\kappa-2J\Delta\right)$, which vanishes at the saddle point by \cref{eq:condcondition}, but its derivatives $\partial f_0/\partial\lambda=n_0$, $\partial f_0/\partial\kappa=4tn_0$, and $\partial f_0/\partial\Delta=-2Jn_0$ do not. They enter the saddle-point equations below as the condensate contributions, since the macroscopic zero-mode occupation is not generated by the continuum integrals (whose singularity at $x=-2$ is integrable in two dimensions) and must be added separately as $n_0$. Eliminating $\lambda$ via \cref{eq:condcondition}, the spinon spectrum takes the scaling form
\begin{equation}
    \begin{aligned}
        E^{b}(x) & =J\Delta\, e_r(x)\,,
        \quad
        r\equiv-\frac{t\kappa}{J\Delta}>0\,,
        \\
        e_r(x)   & =\sqrt{4\left[1+r(x+2)\right]^{2}-x^{2}}\,,
    \end{aligned}
    \label{eq:erdef}
\end{equation}
with $e_r(-2)=0$ corresponding to the condensate mode at $Q$, and the continuum parts of the spinon equations are captured by the three integrals
\begin{equation}
    \begin{aligned}
        I_0(r) & =\int_{-2}^{2}dx\,\rho(x)\,\frac{2\left[1+r(x+2)\right]}{e_r(x)}\,,     \\
        I_1(r) & =\int_{-2}^{2}dx\,\rho(x)\, x\,\frac{2\left[1+r(x+2)\right]}{e_r(x)}\,, \\
        I_2(r) & =\int_{-2}^{2}dx\,\rho(x)\,\frac{x^{2}}{e_r(x)}\,.
    \end{aligned}
    \label{eq:Iints}
\end{equation}
The zero-temperature spinon equations \eqref{eq:saddlelambda}, \eqref{eq:saddlekappa}, and \eqref{eq:saddledelta} then become
\begin{equation}
    \begin{aligned}
        I_0(r)+n_0          & =2-\delta\,, \\
        I_1(r)+4\chi-2n_0   & =0\,,        \\
        2\Delta-I_2(r)-2n_0 & =0\,.
    \end{aligned}
    \label{eq:saddleT0}
\end{equation}
Since the holon sector has already fixed $\kappa=\kappa_\delta$, the definition of $r$ gives $\Delta=K_\delta/r$ with $K_\delta\equiv-t\kappa_\delta/J>0$; combining it with the first and third relations of \cref{eq:saddleT0} yields the single scalar equation $K_\delta/r=\tfrac{1}{2}I_2(r)+2-\delta-I_0(r)$, whose solution $r=r_\delta$ determines the condensed spinon density
\begin{equation}
    n_0(\delta)=2-\delta-I_0(r_\delta)\,.
    \label{eq:n0delta}
\end{equation}

For $t=1$ and $t/J=2$, the resulting staggered moment $m_s=n_0/2$ is shown in \cref{fig:mft-ms} of the main text. The condensate density $n_0$ rises from $n_0(0.02)\approx 0.7095$ to its maximum $n_0^{\max}\approx 0.7243$ near $\delta\simeq 0.053$ and then decreases slowly to $n_0(0.125)\approx 0.6991$, while the condensate \emph{fraction} $n_0/(1-\delta)$ increases monotonically over this range, since the total spinon density $1-\delta$ decreases faster than $n_0$. As $\delta\rightarrow 0$ the moment smoothly approaches the Schwinger-boson mean-field benchmark $m_s\simeq 0.3034$ of the half-filled Heisenberg antiferromagnet \cite{arovas1988functional,sarker1989bosonic}. Dilute doping \emph{enhances} the in-plane antiferromagnetic order, through the kinetic mechanism discussed in \cref{sec:mft} and \cref{sec:semiclassical}.

\subsection{Mean-field theory of the \texorpdfstring{$\mathbb{Z}_2(0,\pi)$}{Z2(0,pi)} state}
\label{app:mf0pi}

Finally, we briefly present the mean-field theory of the other allowed zero-flux state, $\mathbb{Z}_2(0,\pi)$. Substituting the labels of \cref{eq:labels0} into \cref{eq:psgsolution}, the PSG differs from \cref{eq:psg00} only through $G_d(x,y)=i$. Applying the transformation rules \cref{eq:chitrans,eq:deltatrans} to the representative $x$-bond amplitudes of \cref{eq:fiducial}, the hopping and all translation-related bonds are unchanged, while $\braket{\hat{\Delta}_{i,i+\hat{y}}} = -\braket{\hat{\Delta}_{i,i+\hat{x}}} = - i\Delta$, giving the pairing a $d$-wave-like structure. Repeating the mean-field construction of \cref{sec:mf-solution} with this bond pattern, the holon sector is unchanged, while the spinon pairing form factor is replaced by $\mathcal{E}_k\rightarrow\widetilde{\mathcal{E}}_k$, so that the Bogoliubov-spinon dispersion becomes
\begin{equation}
    \begin{aligned}
        E^{b}_{k}                 & =\sqrt{\left(\lambda-2t\kappa\,\mathcal{E}_k\right)^{2}-J^{2}\Delta^{2}\,\widetilde{\mathcal{E}}_k^{2}}\,, \\
        \widetilde{\mathcal{E}}_k & =\sin k_x-\sin k_y\,.
    \end{aligned}
    \label{eq:Eb0pi}
\end{equation}
The self-consistency equations retain exactly the form of \cref{eq:saddlelambda,eq:saddlemu,eq:saddlekappa,eq:saddlechi,eq:saddledelta}, with $E^{b}_{k}$ given by \cref{eq:Eb0pi} and with $\mathcal{E}_k^{2}\rightarrow\widetilde{\mathcal{E}}_k^{2}$ in the gap equation \eqref{eq:saddledelta}; in the undoped limit the spinon hopping vanishes together with $\kappa$, and the two states reduce to the same case, their spectra being related by $k_y\rightarrow -k_y$.

The magnetic order upon spinon condensation follows as in \cref{sec:mf-solution}, with one crucial difference. The pairing form factor now vanishes at the condensation momenta, $\widetilde{\mathcal{E}}_{\pm Q}=0$, so the soft Nambu sector $(b_{Q\uparrow},\, b^{\dagger}_{-Q\downarrow})$ is purely diagonal, with both modes at the energy $\lambda+4t\kappa$. In contrast with \cref{eq:Hbsoft}, the pairing cannot lock the two components, and when the gap closes they condense with \emph{independent} amplitudes, $\braket{b_{Q\uparrow}}=\sqrt{N}\,z$ and $\braket{b_{-Q\downarrow}}=\sqrt{N}\,w$. The condensate amplitudes $\braket{b_{i\uparrow}}=z\,e^{-iQ\cdot r_i}$ and $\braket{b_{i\downarrow}}=w\,e^{iQ\cdot r_i}$ then yield the local magnetization
\begin{equation}
    \braket{S^{z}_i}=\frac{1}{2}\left(|z|^{2}-|w|^{2}\right),
    \quad
    \braket{S^{+}_i}\propto(-1)^{x_i+y_i}\, w z^{*}\,.
    \label{eq:order0pi}
\end{equation}
The condensation in the $\mathbb{Z}_2(0,\pi)$ state therefore produces antiferromagnetic order in the $xy$ plane \emph{together with} a uniform ferromagnetic moment along the $z$ axis, since generically $|z|\neq|w|$. This provides a further, qualitative distinction from the $\mathbb{Z}_2(0,0)$ state, whose condensate carries no $z$-axis moment, as given by \cref{eq:spinorder}. Both the lower mean-field energy and the absence of a spontaneous $z$-axis magnetization in the iPEPS ground state study \cite{zheng2025revealing} thus consistently select the $\mathbb{Z}_2(0,0)$ state as the mean-field description of the $\sigma t$-$J$ model.

\bibliography{refs}

\end{document}